\documentclass[aps,pra,10pt,superscriptaddress,reprint]{revtex4-1}

\usepackage{amstext}
\usepackage{fullpage}
\usepackage{amssymb,amsmath} 
\usepackage{amsthm} 
\usepackage{bbold}
\usepackage{hyperref}
\usepackage{graphicx} 
\usepackage{multirow} 
\usepackage[table]{xcolor}

\newcommand{\be}{\begin{equation}}\newcommand{\ee}{\end{equation}}
\newcommand{\ba}{\begin{eqnarray}}\newcommand{\ea}{\end{eqnarray}}
\newcommand{\ban}{\begin{eqnarray*}}\newcommand{\ean}{\end{eqnarray*}}

\newcommand{\ket}[1]{\mbox{$ | #1 \rangle $}}
\newcommand{\bra}[1]{\mbox{$ \langle #1 | $}}
\newcommand{\avg}[1]{\langle #1 \rangle}

\DeclareMathOperator{\tr}{Tr}

\begin{document}

\title{Quantum randomness extraction for various levels of characterization of the devices}


\author{Yun Zhi Law}
\affiliation{Centre for Quantum Technologies, National University of Singapore, 3 Science Drive 2, Singapore 117543, Singapore}
\author{Le Phuc Thinh}
\affiliation{Centre for Quantum Technologies, National University of Singapore, 3 Science Drive 2, Singapore 117543, Singapore}
\author{Jean-Daniel Bancal}
\affiliation{Centre for Quantum Technologies, National University of Singapore, 3 Science Drive 2, Singapore 117543, Singapore}
\author{Valerio Scarani}
\affiliation{Centre for Quantum Technologies, National University of Singapore, 3 Science Drive 2, Singapore 117543, Singapore}
\affiliation{Department of Physics, National University of Singapore, 2 Science Drive 3, Singapore 117542, Singapore}

\date{\today}

\begin{abstract}
The amount of intrinsic randomness that can be extracted from measurement on quantum systems depends on several factors: notably, the power given to the adversary and the level of characterization of the devices of the authorized partners. After presenting a systematic introduction to these notions, in this paper we work in the class of least adversarial power, which is relevant for assessing setups operated by trusted experimentalists, and compare three levels of characterization of the devices. Many recent studies have focused on the so-called ``device-independent" level, in which a lower bound on the amount of intrinsic randomness can be certified without any characterization. The other extreme is the case when all the devices are fully characterized: this ``tomographic" level has been known for a long time. We present for this case a systematic and efficient approach to quantifying the amount of intrinsic randomness, and show that setups involving ancillas (POVMs, pointer measurements) may not be interesting here, insofar as one may extract randomness from the ancilla rather than from the system under study. Finally, we study how much randomness can be obtained in presence of an intermediate level of characterization related to the task of ``steering", in which Bob's device is fully characterized while Alice's is a black box. We obtain our results here by adapting the NPA hierarchy of semidefinite programs to the steering scenario.
\end{abstract}

\maketitle

\section{Introduction}

The quantum information community has always duly acknowledged Bell's work as pioneering. Indeed, Bell's discovery \cite{bell1964} that an apparently philosophical issue can actually be settled by experiment was a precursor to the approach of stopping complaining about quantum weirdness and putting it to practical use. Once this novel approach was adopted, however, there seemed to be little scope left for Bell inequalities: Bell having done its job by showing that local variables don't exist, one should focus on entanglement as a resource. A few authors tried to find links between Bell inequalities and some useful tasks, but the reported examples happened to be either artefacts of restrictive assumptions \cite{SG02} or \textit{ad hoc} constructions \cite{vienna04}. 

The practical role of Bell inequalities in quantum information was fully clarified around the year 2007 (see \cite{review,slovaca} for a review of those developments): the violation of Bell inequalities witnesses entanglement without any need to specify the dimensionality of the system under study or the measurements that are performed on it. This means that one can certify, and even more, quantify, the entanglement shared between two or more \textit{black boxes}. This possibility, unique to Bell inequalities, has been called \textit{device-independent certification}. The first task to be studied this way was quantum key distribution \cite{acin2007device}; it was followed by the task of generating \textit{certified randomness} \cite{pironio10random,colbeck11private} and by the quantifications of some entanglement measures~\cite{bardyn09,moroder2013}.

Device-independent certification still requires some assumptions on the devices, that we are going to describe in detail in the next section: these assumptions define the minimal \textit{level of characterization} of the devices that is required to be able to certify some quantum behavior. On the contrary, the bulk of the quantum information literature usually works under the assumption that the degrees of freedom under study (``system") and the measurements are fully characterized --- intriguingly, this is the case even in quantum cryptography: for instance, the unconditional security of the BB84 protocol relies on the fact that the systems used by Alice and Bob are qubits \cite{agm06}. Between the two extremes, several semi-device-independent levels of trust can be relevant.

The goal of this paper is to discuss how the production of \textit{quantum randomness} depends on the level with which the devices involved are characterized. Our results will be derived in the minimal adversarial class which is relevant to assess setups produced and operated by honest experimentalists. We begin by clarifying all these scenarios (Section \ref{sec2}), then we introduce the mathematical tools (Section \ref{sec_DefAndNot}) and present an explicit comparison between three levels of characterization (Section \ref{sec_compare}). Finally, we highlight some subtleties proper to randomness extraction in a scenario of complete characterization of the system and the measurements (Section \ref{sec_more}).

\section{Scenarios for quantum randomness}
\label{sec2}

\begin{figure}[!h]
	\centering
		\includegraphics[width=0.2\textwidth]{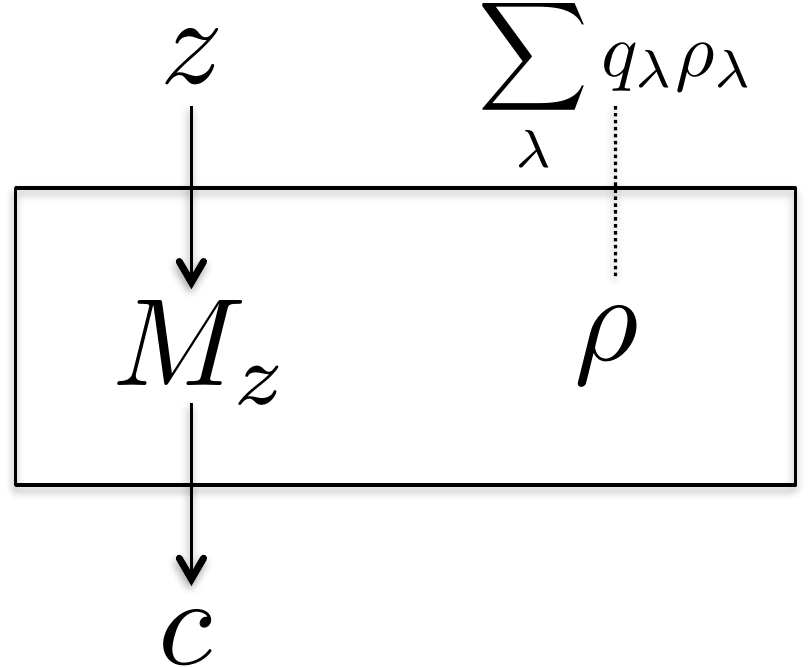}
	\caption{Generic setup of quantum randomness extraction. The authorized party inputs $z$ into a box and receives the outcome $c$, whose randomness one wants to guarantee with respect to an adversary Eve. Inside the box, the role of $z$ consists in selecting a possible measurement $M_z$ to be performed on a state $\rho$, and $c$ is the outcome of that measurement. The assumption of \textit{measurement independence} (which is enforced here, though it could be partially relaxed) means that in each run $z$ and $\rho$ must be uncorrelated. Various scenarios can be considered based on the power given to Eve (discussed in paragraph \ref{sseve}) and the level of characterization that the authorized party has of her devices (paragraph \ref{sstrust}).}
	\label{fig_gensetup}
\end{figure}

We start this paper by a concise review of various scenarios for quantum randomness. The generic setup for quantum randomness is sketched in Fig.~\ref{fig_gensetup}. As explained in the caption, the goal of the authorized party is to certify the randomness of the outcome $c$ \textit{with respect to Eve's knowledge}. There is \textit{a priori} place for a third party, the provider, who may have produced some of the devices but has no interest in learning $c$. A scenario will be defined by the power given to Eve and the level of characterization of the devices. We address them in this order.

For definiteness, we focus exclusively on the case of \textit{measurement independence}: namely, we assume that the state to be measured $\rho$ and the choice of the measurement $z$ are fully uncorrelated in each run. It is remarkable that even this assumption can be partially relaxed, giving rise to the possibility of randomness amplification~\cite{colbeck12amplifysv,grudka13amplifysv,gallego13amplifysv,ramanathan13amplifysv,brandao13amplifysv,mironowicz13amplifysv,bouda14amplifymin,chung14amplifymin}, but we don't consider this task in this paper.

\subsection{Classes of adversarial power}
\label{sseve}

Throughout the paper, we are going to assume that quantum theory holds; in particular, we don't discuss the possibility of certifying randomness against an adversary limited only by no-signaling~\cite{acin06,colbeck12amplifysv,gallego13amplifysv,ramanathan13amplifysv,brandao13amplifysv,coudron13expand}. The power given to Eve can be divided in three main classes:
\begin{itemize}
\item[Class (I)] Eve has \textit{no access to the devices}. Because of the Kerchhoff-Shannon principle (one should not look for security in hiding details of the hardware or the protocol), we assume that even though she cannot influence it, \textit{Eve has knowledge of the experimental setup}. Thus, she may know in each run which quantum state enters the box, and which measurement is used. Her description of both state and measurements may be better than that of both Alice and the provider: for instance, she may be able to describe the state as pure in each run by knowing which decomposition of a possible mixture is being used. 
Since the choice of measurement in each run is known to Eve, no initial private randomness is required here.
We can define two subclasses, according to whether Eve does not (a) or does (b) hold a purification of the state in a quantum memory (in the literature on quantum cryptography one finds also intermediate situations between (a) and (b), the so-called bounded-storage models~\cite{damgard08} and noisy storage models~\cite{wehner08noisy}). Because quantum theory is no-signaling, holding a purification does not allow Eve to change the state $\rho$, but may give her more guessing power since she will be able to steer towards a specific decomposition (by performing a specific measurement on her reduced state).

\item[Class (II)] Eve \textit{distributes the state} to the users, but has only classical, though perfect, knowledge about the set of possible measurements. Note that in this case, the measurement settings should not be known by Eve in advance, otherwise she could prepare the state accordingly to obtain full knowledge of the outcomes. Thus, an initial amount of private randomness is necessary to run the protocol.
We can again define subclasses (a) and (b) as above. Class II(b) is a natural analog of the entanglement-based scheme in which quantum cryptography can be proven ``unconditional secure": as such, most papers on quantum randomness have considered it~\cite{vazirani11expand,coudron13expand,coudron13,miller14expand}. This class is natural in the case where Alice holds two subsystems situated in secured, but possibly distant locations.

\item[Class (III)] The maximal class is that in which \textit{Eve is the provider}, namely she prepares both the measurement devices and the state: the only power left with Alice is the choice of $z$ in each run. The claim is frequently made, at least in semi-popular accounts, that quantum physics could provide security against an adversarial provider. The correct statement is that one may be able to \textit{certify the quantumness} of the process that generates $c$, typically in the case of a loophole-free Bell test. However, an adversarial provider would certainly find a way to hide a transmitter in the devices, with the task of leaking out the values of $c$ at the end of the protocol, or may employ attacks discussed in~\cite{barrett13attackdiqkd}. Therefore, however certifiably quantum the process that generated the outcomes may have been, there cannot be any randomness with respect to an adversarial provider. So we won't consider this class any longer in this paper.
\end{itemize}

A point worth stressing is the amount of randomness needed to generate the seed $z$. The choice of this seed must indeed guarantee measurement independence, i.e. $z$ must be ``random" with respect to the choice of the state in each run, and viceversa; while the randomness we want to extract must be unpredictable for Eve. Therefore, if one works in Class I, no randomness with respect to Eve is required in the input seed: even if Eve knows $z$, she cannot adapt the state to it. Similarly, the random seed needed to extract the final random bits from the output $c$ through the left-over hash lemma needs only be random with respect to the outputs themselves. It can thus be known to the adversary. One can then speak of \textit{randomness generation}~\cite{pironio13expand}. On the contrary, in Classe II, since Eve prepares the state, the inputs $z$ must be generated with a process that is \textit{a priori} guaranteed to be unknown to Eve. In other words, in these classes there is no randomness generation, but only \textit{randomness expansion} --- and, after a series of partial results~\cite{vazirani11expand,fehr13expand,pironio13expand}, it has been proved that such expansion can in principle be unbounded~\cite{coudron13expand}. The different adversarial classes discussed here are summarized in Table~\ref{tab:classes}.

\begin{widetext}
\begin{center}

\begin{table}
\centering
\begin{tabular}{c|c|c|c|}
& \normalsize Class I & \normalsize Class II & \normalsize Class III \\
\textit{Eve...} & \textit{has no access to the devices} & \textit{distributes the state} & \textit{prepares the devices} \\ \hline & \cellcolor{black!25} & & \\[-5pt]
\normalsize Subclass (a) &  \cellcolor{black!25} \normalsize randomness &  &  \\
\textit{holds classical side information} & \cellcolor{black!25} \normalsize generation & \normalsize randomness & \normalsize no\\ \cline{1-2} & & \normalsize expansion & \normalsize randomness\\[-5pt]
\normalsize Subclass (b) & \normalsize randomness & & \\
\textit{holds quantum side information} & \normalsize generation & & \\ \hline
\end{tabular}
\caption{Randomness protocols in presence of measurement independence with different classes of adversaries. The gray cell corresponds to the trusted provider assumption, which describes the class of adversaries considered in this work.}\label{tab:classes}
\end{table}

\end{center}
\end{widetext}

Finally recall that we are assuming strict measurement independence. If partial measurement dependence is taken into account, one would have to refine the definition of these classes: for instance, by specifying whether the dependence is introduced unwittingly by the provider or maliciously by Eve.

\subsection{Levels of characterization of the devices}
\label{sstrust}

The second defining feature of a scenario is the level of characterization that Alice has of the working of her devices. We sketched it in the introduction, now we can be more precise:
\begin{itemize}
\item The \textit{tomography} level of characterization usually means that Alice knows the behavior of her devices as well as Eve does. This is what we shall consider in this paper. Of course, in the vast literature on quantum tomography, the possibility of partial knowledge of the devices has been considered~\cite{james01,moroder10,teo12,rosset12}, so one could refine this level of characterization in several sublevels. In all cases, though, it is assumed that Alice knows exactly which degrees of freedom are relevant to the measurement (polarization, spin, quadrature of a field...).

When this trust in the characterization is unwarranted, a Class II Eve may successfully \textit{hack} the devices. In particular, Eve may know that some degrees of freedom, others than the ones Alice is aware of, play a role in the physical process, and may thus influence the behavior of the boxes by addressing those degrees of freedom. For instance, this was the case in the series of experiments that hacked quantum cryptography devices by exploiting the physics of photodetectors~\cite{lydersen10qkdhack,liu14qkdhack,tanner14qkdhack}. Similarly, a Class I Eve could also take advantage of her knowledge about what happens in additional degrees of freedom if a setup happens to use them.

\item The \textit{device-independent} level of characterization means that Alice does not rely on a description of her devices but only on the observed statistics. Since the statistics on a single quantum system can always be reproduced with classical randomness, device-independence requires a loophole-free violation of a Bell inequality. Specifically, it is crucial to close the detection loophole, while the no-signaling condition may be justified in other ways than by arranging spacelike separation. In any case, Alice may actually hold both measurement devices in her lab. Nevertheless, for convenience we shall speak of Alice and Bob when it comes to device-independence. We also note that, as it happens for measurement independence, the strict no-signaling condition may be partially relaxed, but we don't consider this relaxation in this paper~\cite{hall11relaxed}.

\item One can define intermediate levels of characterization, collectively known as \textit{semi-device-independent}. For instance, in the context of quantum cryptography, the idea of \textit{measurement-device-independence} has been put forward after noting that hacking usually involves detectors (which are therefore better left untrusted) rather than sources (which may therefore be trusted) \cite{measdi}. Other works relax the tomographic requirement of perfect knowledge of the degree of freedom, but assume an upper bound on the dimensionality of the systems under study \cite{Marcin2011}. In this paper, we shall consider explicitly the case called \textit{one-sided device-independent}, in which entanglement is certified by two devices, Alice's being tomographic and Bob's unknown; this case was studied for quantum key distribution \cite{steer12}.
\end{itemize}

\subsection{Assumptions of this paper}

In this paper, we consider Eve in \textit{Class I(a)}, which was called ``trusted provider" in previous works \cite{bancal13more}. In real life, this class describes randomness generation in an experiment performed by a trusted laboratory: therefore, even if it does not explore the ultimate limits of quantum power, it is arguably relevant for physics~\cite{giustina12,kwiat13exp}. The randomness is guaranteed against any Eve that is not involved in setting up or running the experiment. Furthermore, Eve not being in the lab, she could hold the purification only if those degrees of freedom were ``radiative" and she had the power to collect them: this, together with the state-of-the-art of quantum memories, makes it very reasonable to restrict to subclass (a) at least for a few years to come.

Also, we consider only the asymptotic limit of infinitely many runs where each run is assumed to be independent and identically distributed (i.i.d.) according to some strategy of Eve. Corrections due to finite samples, and extension to non-i.i.d. scenarios can in principle be done with the techniques in \cite{gill03,knill11,pironio13expand}.

\section{Computing randomness for different levels of characterization}
\label{sec_DefAndNot}

\subsection{Definitions and notation}

Let us introduce the basic notions to study randomness (cf.~\cite{acin12randomness}). The authorized party can input $z\in\{1,...,m\}$ in the box and obtain output $c\in\{1,...,d\}$ (see Fig.~\ref{fig_gensetup}). In the asymptotic limit of infinitely many runs, she can reconstruct the statistics $P(c|z)$. This statistical distribution may reflect either accidental randomness (due to ignorance of some details of the state or the device) or intrinsic randomness, due to the unpredictability of the outcome of quantum measurements. We are interested in the latter because for Eve, in any of the Classes defined above, there is no accidental randomness.

In this section, the states $\ket{\psi}$ or $\rho$ refer to everything that is in the box, so that the measurements can be assumed to be projective. When the box is fully characterized, it becomes possible to distinguish between the system and a possible ancilla, i.e. to discuss the case of POVMs as distinct. We do this in section~\ref{sec_povm}.

If the state is pure, there is no accidental randomness for von Neumann measurements. Then, the randomness of $c$ obtained from a given $z$ is quantified by the probability $G(\ket{\psi},z)$ of guessing the outcome correctly. Since the best strategy for guessing is to guess the most probable outcome, this guessing probability is
\ba
	G(\ket{\psi},z) &=& \max_{c} P(c| z,\ket{\psi}) \label{eq_guessing}.
\ea
If the state shared by Alice and Bob is mixed, we have to separate the intrinsic randomness from the accidental one. For measurement $z$, the average guessing probability that quantifies intrinsic randomness is then given by
\begin{eqnarray}\label{guessmixed}
	G(\rho,z) &=& \max_{ \{q_\lambda, \psi_{\lambda} \} } \sum_{\lambda} q_{\lambda} G(\ket{\psi_{\lambda}}, z) \label{eq_guessing_mixed_state},
\end{eqnarray}
where $\rho = \sum_{\lambda} q_{\lambda} \ket{\psi_{\lambda}}\bra{\psi_\lambda}$, and the maximization is taken over all possible such decompositions.

For the device-independent level of characterization, Alice cannot write down a quantum state but needs to use only the observed probability distribution $P$. Then the guessing probability that quantifies intrinsic randomness is
\begin{eqnarray}
	G(P,z) &=& \max_{ ( \rho,M ) \rightarrow P } G(\rho,z) \label{eq_guessing_black_box},
\end{eqnarray}
where the maximization is taken over all quantum states $\rho$ and measurements $M$ compatible with the probability distribution $P$.

In all these cases, the number of random bits that can be extracted per run is quantified by the min-entropy
\begin{eqnarray}
	H_{\min}(G) &=& - \log_2 G\,. \label{eq_defineGlobalRandom}
\end{eqnarray}

Finally, one may consider extracting randomness out of several settings, rather than a single one. If Eve is allowed to keep a purification of the state [subclasses (b)], upon learning which settings have been used in a given run, she may be able to steer the state to the decomposition that maximizes \eqref{guessmixed} for those settings~\cite{footnote}. 
If Eve does not hold a purification [subclasses (a)], however, using several settings is known to be advantageous \cite{bancal13more}.

Now we are going to explain how randomness can be computed in each of the three levels of characterization of our concern. For the device-independent level of characterization, randomness generation against a Class Ia Eve have been presented in~\cite{silleras14more,bancal13more}, so we don't repeat this here.

\subsection{Tomography level of characterization}
\label{sec_tomo}

For the case of tomography level of characterization, we are on a ground familiar for most physicists. If a qubit is prepared in the state $\ket{+z}$, to say that a measurement of $\sigma_x$ provides a perfect random bit is just a rephrasing of elementary textbook knowledge. The example of a qubit prepared in the maximally mixed state $\mathbb{1}/2$ is only slightly more involved: then, a measurement of a single observable (say) $\sigma_x$ does not guarantee any randomness, because the state may have been prepared by mixing eigenstates of that operator, in which case Eve would have full knowledge of the outcome of each run. However, the uncertainty relations provide a way around it: if in each run Alice can choose to measure either $\sigma_x$ or $\sigma_z$, no preparation can be an eigenstate of both, therefore there is randomness with respect to Eve as long as she does not hold a purification (see~\cite{tomamichel11} for how uncertainty relations must be modified if Eve does hold a purification).

Now we provide a general recipe to compute the intrinsic randomness for projective measurements; the case of POVMs will be discussed in paragraph \ref{sec_povm}.

Since the state $\rho$ can be reconstructed and is therefore part of the observed data, we need to perform the maximization of Eq.~\eqref{eq_guessing_mixed_state}. In the decomposition, it is not \textit{a priori} obvious how many quantum states $\ket{\psi_\lambda}$ are to be considered. Fortunately, the argument used in~\cite{bancal13more} can be transposed directly from probability distributions to density matrices. In a nutshell, all the lambdas for which the inner maximization is achieved by the same argument $c$ can be grouped together into a unique lambda (which is relabeled as $c$). Thus, it is sufficient to consider \textit{one state per outcome}. 

Therefore, for a projective measurement $M\equiv\{\Pi_c, c=1,...,d\}$ with $d$ outcomes, one has to solve
\begin{eqnarray}
G(\rho, M)&=& \max_{\{\rho_{c}\}} \sum_{c} \tr\big[ \rho_{c} \Pi_{c} \big] \label{eq_guessing_tomo_1gen}
\end{eqnarray} under the constraints that $\rho = \sum_{c}  \rho_{c}$, $\rho_{c} \geq 0$. Like in the case of device-independence, this maximization is a semi-definite program (SDP), the only difference being that the matrix that must be positive is the quantum state itself, not a matrix of momenta of the observed statistics. Moreover, here the SDP solves the problem of interest directly, rather than a relaxation thereof. Given the tomography level of characterization, Alice can choose to extract randomness from \textit{any} measurement, and will choose that for which the guessing probability in \eqref{eq_guessing_tomo_1gen} is the lowest. Hence, for a given state $\rho$,
\begin{eqnarray}
	G(\rho)=\min_{M} G(\rho, M) \textrm{ [one measurement].}\label{eq_guessing_tomo_2}
\end{eqnarray}
Further, when Eve is not allowed to hold a purification, it may be advantageous to extract randomness from more measurements. If setting $M_z$ is chosen with probability $q_z$, the average guessing probability will be
\begin{eqnarray}
G(\rho, \{M_z\}_z)&=& \max_{\{\rho_{C}\}} \sum_{C} \sum_{z=1}^m q_z \tr\big[ \rho_{C} \Pi^z_{c_z} \big]\nonumber\\
&=& \max_{\{\rho_{C}\}} \sum_{C} \tr\big[ \rho_{C} {\cal M}_C \big] \label{eq_guessing_tomo_2gen},
\end{eqnarray}
where we have denoted ${\cal M}_C=\sum_{z=1}^m q_z\Pi^z_{c_z}$; the constraints are as above, and now $C=(c_1,c_2,...,c_m)$, so the maximization now involves a decomposition on $d^m$ states. The fact that Eve cannot steer Alice's mixture is explicit in that the decomposition $\rho=\sum_C \rho_C$ is independent of $z$. As above, Alice is allowed to choose the set of measurements that minimizes the guessing probability, so
\begin{eqnarray}
	G(\rho) &=& \min_{\{M_z\}} G(\rho, \{M_z\}) \textrm{ [more measurements].} \label{eq_guessing_tomo_3}
\end{eqnarray}
Notice that unlike the optimization (\ref{eq_guessing_tomo_2gen}) which is an SDP, this last optimization over the choices of measurement settings is not a SDP.

\subsection{One-sided device-independent level of characterization} \label{sec_semi_device_independent}

The one-sided device-independent level of characterization, to our knowledge, has never been considered before in the context of randomness. The scenario is very similar to steering \cite{jones2007entanglement, cavalcanti2009experimental}: the setup is actually the same, but the figure of merit is different. Indeed, instead of having Alice to convince Bob that she can steer his state, we just let them perform their measurements locally and ask whether randomness can be extracted from their outcomes.

Like before, we consider first the amount of random bits that can be extracted from the outcomes $c = (a,b)$ of a single pair of measurements  $z = (x,y)$ with $A_x=\{\Pi^x_a\}$ and $B_y=\{\Pi^y_b\}$ denoting Alice and Bob's local measurements. The guessing probability is analog to \eqref{eq_guessing_black_box} and given by
\begin{eqnarray}
	G(P, z) &=& \max_{ ( \rho,A_x,B_y ) \rightarrow P } G(\rho,A_x,B_y) \nonumber \\
	&=& \max_{ ( \rho_{c},A_x,B_y ) \rightarrow P } \sum_{c} P_{c}(c|z) \label{eq_steering_optimize_problem}
\end{eqnarray}
where $\rho  = \sum_{c} \rho_{c}$, and $P_{c}(a,b|x,y) = \tr(\rho_{c} \Pi^x_a \otimes \Pi^y_b)$. The constraints for the optimization are the observed statistics $P(a,b|x,y)$, and the knowledge of the state and measurements on Bob's side.

Such optimization is very similar to the one used for the device-independent level of characterization in~\cite{bancal13more,silleras14more}, where one can use the hierarchy introduced in~\cite{qtest} to provide upper bounds. In that case, from the set of local measurements and depending on the hierarchy's level, one forms a certain matrix $\Gamma_{c}$  whose elements are expectation values with $\rho_c$ of products of operators of the form: $\avg{M_A\otimes\mathbb{1}}$, $\avg{\mathbb{1}\otimes M_B}$, $\avg{M_A\otimes M_B}$, $\avg{M_AM_A'\otimes\mathbb{1}}$, $\avg{\mathbb{1}\otimes M_BM_B'}$, $\avg{M_AM_A'\otimes M_BM_B'}$, etc, where $M_A,M_A'$, and $M_B,M_B'$ are operators from the set of Alice's and Bob's local measurements (union the identity), respectively (see~\cite{qtest} for a detailed description of this matrix). Some elements of $\Gamma_c$ are related to the $P_{c}(a,b|x,y)$ mentioned above, while others are extra unknown variables in the optimization. By constraining $\Gamma_{c}$ to be positive semi-definite and the sum over $c$ of $P_{c}(a,b|x,y)$ to be the observed statistics, one can bound the guessing probability in the device-independent level of characterization.

Now in the one-sided device-independent case, we impose further constraints on the elements of $\Gamma_c$ based on the knowledge of Bob's measurements. Namely, we use the algebraic relations satisfied by these operators to constraint the moments of $\Gamma_c$ which involve them. For instance, if $B_3 = (B_1 + B_2)/\sqrt{2}$ or $B_1B_2=-B_2B_1$, the relations  $\avg{\mathcal{O} B_3} = (\avg{\mathcal{O}B_1} + \avg{\mathcal{O} B_2})/\sqrt{2}$ or  $\avg{\mathcal{O}B_1B_2}=-\avg{\mathcal{O}B_2B_1}$ are imposed for all product of operators $O$. This reduces the number of independent variables in the optimization. These relations are imposed on each $c$ in~\eqref{eq_steering_optimize_problem}, as they should hold for each of the $\rho_c$ in the decomposition. Note that we do not directly use the knowledge of Bob's local state to further constraint the optimization. We thus obtain an upper bound on the guessing probability.

Just like in the other levels of characterization, one could consider extracting randomness from more than one measurement here, if Eve does not hold a purification of the measured state. The optimization problem can be set up in a manner analogous to what we have been considering.

\section{Comparison of the yields of three levels of characterization} \label{sec_compare}

In order to compare the yields of the various levels of characterization, we need a common set of data. We assume that the data come from measuring a two-qubit Werner state $\rho_V=V\ket{\Phi^+}\bra{\Phi^+}+(1-V)\mathbb{1}/4$; Alice measures either $A_1=\sigma_x$ or $A_2=\sigma_z$, Bob measures one of the four $B_1=\sigma_x$, $B_2=\sigma_z$, $B_3=\sigma_+$ or $B_4=\sigma_-$ with $\sigma_{\pm}=(\sigma_x\pm\sigma_z)/\sqrt{2}$. These measurements can lead to non-trivial assessment for all the level of characterization we are interested in. Indeed, $(A_1,A_2;B_1,B_2)$ can be used for partial tomography and identify $\ket{\Phi^+}$ uniquely when $V=1$; for device-independence, $(A_1,A_2;B_3,B_4)$ violate the CHSH inequality for $V>1/\sqrt{2}$ and certify $\ket{\Phi^+}$ for $V=1$ because $\textrm{CHSH}=2\sqrt{2}$ \cite{popescu92states}; for one-sided device-independence, a similar argument holds for $(A_1,A_2;B_1,B_2)$ and the steering inequality $S_2$ defined in~\cite{saunders10exp}. Anyway, in what follows, the amount of randomness is computed directly from the observed statistics, without processing them into a specific tomography protocol or inequality.

As mentioned before, we focus on the randomness generated by a single pair of setting. In the device-independent case, we bound the amount of random bits using the second level of the hierarchy~\cite{qtest}, as described in \eqref{eq_guessing_black_box} (see~\cite{bancal13more,silleras14more} for more details). For the one-sided device-independent case, we use the method described in section \ref{sec_semi_device_independent} at the same level of the hierarchy, together with the algebraic relations generated by Bob's measurements (including $B_3=(B_1+B_2)/\sqrt{2}$, $B_4=(B_1-B_2)/\sqrt{2}$, $B_1B_2=-B_2B_1$, $B_3B_4=-B_4B_3$, etc.). Finally, for tomography characterization, we compute the amount of randomness based on \eqref{eq_guessing_tomo_2gen}, as explained in section \ref{sec_tomo}.

\begin{figure}[h]
	\centering
		\includegraphics[width=0.5\textwidth]{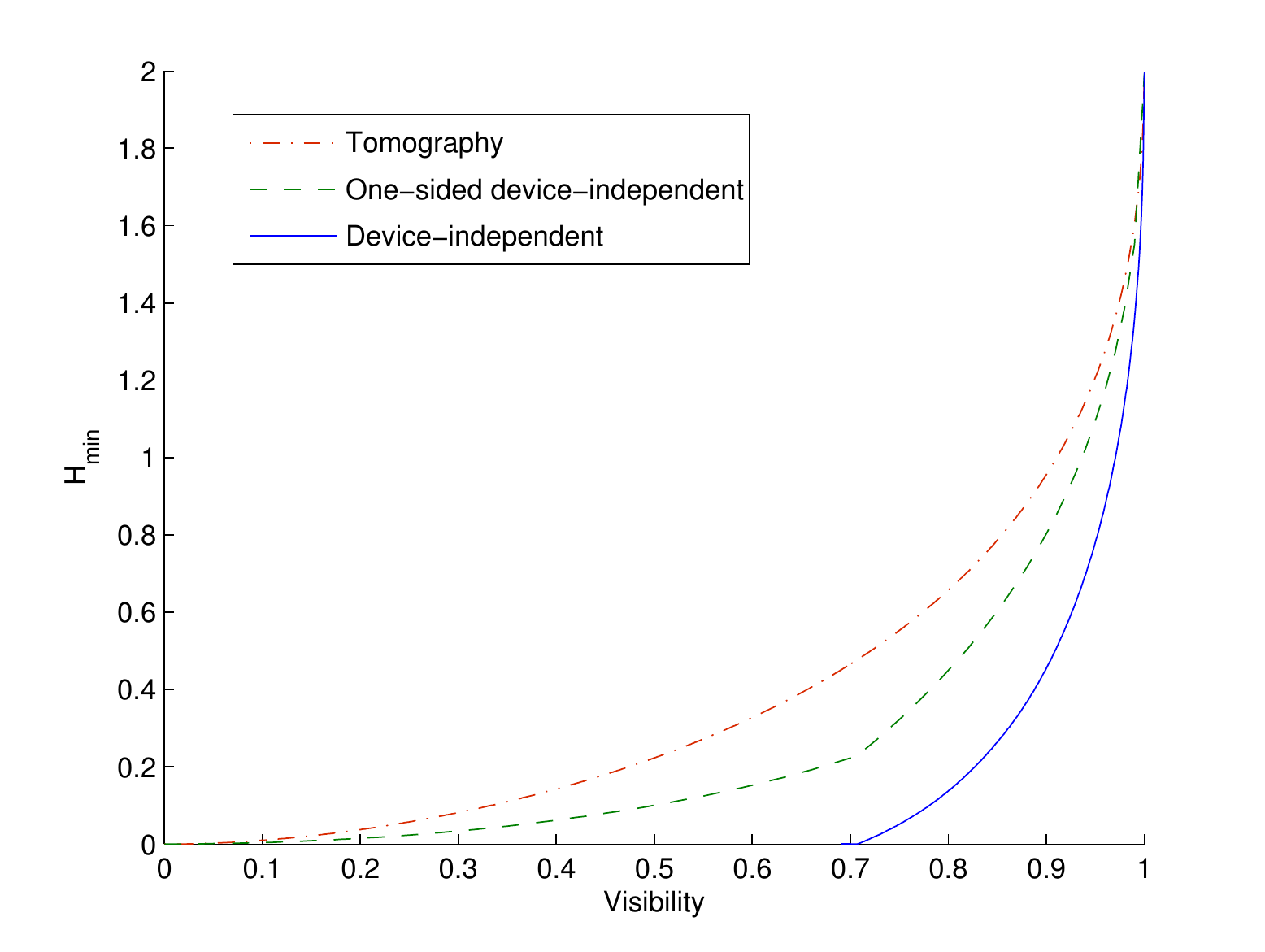}
	\caption{Amount of randomness extracted $H_{\min}$ from the outcomes of the setting pair $A_2,B_1$ from three different levels of characterizations.}
	\label{fig_three_level_comparison}
\end{figure}

The main result we obtained for three different levels of characterization is shown in Fig.~\ref{fig_three_level_comparison}. As expected, for any visibility $V$ the amount of randomness increases with the level of characterization of the devices, with the tomographic level giving the largest amount of randomness.

Note that for all three different levels of characterization, two bits of randomness can be extracted when $V=1$. In the device-independent case, this is more randomness than the $\sim 1.23$ bits that can be certified from the maximal violation of the CHSH inequality~\cite{pironio10random}. To understand this difference, we compare the amount of randomness that can be certified in this scenario from different constraints.

\begin{figure}[h]
	\centering
		\includegraphics[scale=0.3]{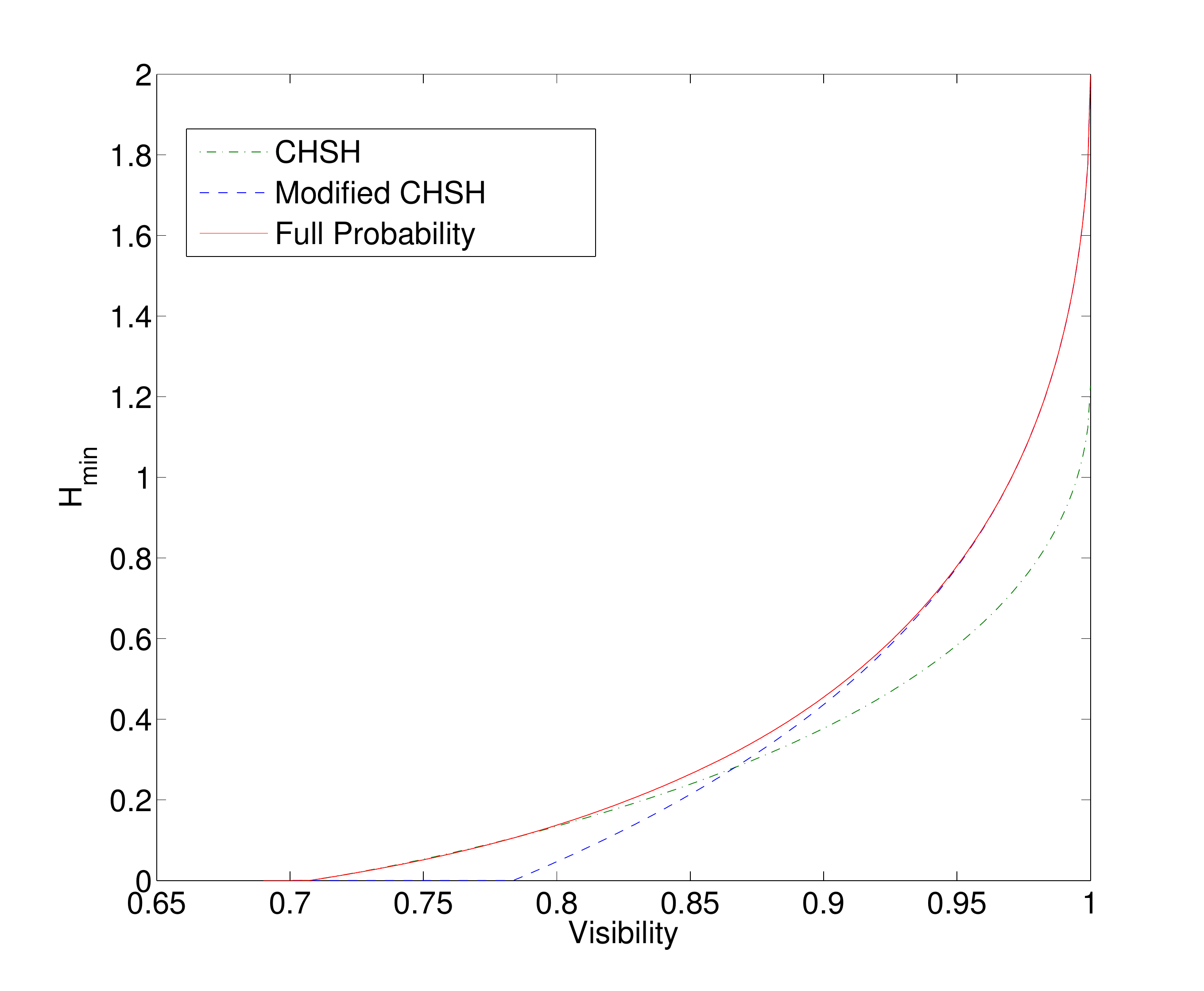}
	\caption{The amount of randomness computed with different constraints: CHSH, $\textrm{CHSH}_3$, and full observed statistics.  $H_{\min}$ of CHSH corresponds to the setting pair $A_2,B_3$, while others correspond to $A_2,B_1$.}
	\label{fig_CHSHvsCHSH3vsConstraintProb}
\end{figure}

Namely, we consider the randomness that can be certified from the correlations above, the one that is certified from an optimal violation of the CHSH inequality
\begin{equation}
	\textrm{CHSH} = \avg{A_1 B_1} + \avg{A_1 B_2} + \avg{A_2 B_1}  - \avg{A_2 B_2} \leq 2\, \label{eq_CHSH}
\end{equation} 
with the same state, and from a optimal violation of a modified CHSH inequality
\begin{eqnarray}
	\textrm{CHSH}_3 &=& \avg{A_1 B_1} + \avg{A_1 B_2} + \avg{A_2 B_1}  \nonumber \\ 
	&& - \avg{A_2 B_2} + \avg{A_1 B_3} \leq 3\,. \label{eq_CHSH3}
\end{eqnarray}  
These last two computations are performed by fixing the value of the inequality rather than the value of the correlations in the corresponding SDP. The result is shown in Fig.~\ref{fig_CHSHvsCHSH3vsConstraintProb}: 2 bits of randomness can be extracted indeed from CHSH$_3$ when the pair of perfectly uncorrelated measurements ${Z_A,X_B}$ are used. However, no such measurements are available when CHSH is maximally violated.

Note also that in Fig.~\ref{fig_three_level_comparison} and~\ref{fig_CHSHvsCHSH3vsConstraintProb}, randomness can be extracted in the device-independent case only provided that a Bell inequality is violated, i.e. when $V < 1/\sqrt{2}$ for CHSH and  $V = 3/(2\sqrt{2}+1) \approx 0.78$ for CHSH$_3$. In other words, no randomness is found when a local hidden variable model can produce the observed correlations.

In the one-sided device-independent case, however, randomness can be extracted from all Werner state, except the completely mixed state. Yet, it is known that Werner states are non-steerable for $V \leq 0.5$~\cite{jones2007entanglement}, i.e. such states admit a Local Hidden State (LHS) model. Thus, our result shows that one can certify randomness in one-sided device-independent context even in presence of a LHS model.

This can be understood by the fact that a local hidden state model only ascribes fixed outcomes to the measurement of one party. The other one, Bob in our case, receives a quantum state to measure. However, Bob enjoys in this context a tomographic level of characterization of his system. He can thus always extract some randomness from this state. This is the case unless all the quantum states given to Bob can all be chosen in the same basis, as possible when $V=0$.

Note that randomness can be extracted for all $V>0$ in the tomographic level of trust as well. However the randomness also disappears there when $V=0$, i.e. when the Werner state is white noise. In the next section, we discuss how randomness can be extracted from the white noise by using several measurement settings.

\section{More results on the tomographic level of characterization}
\label{sec_more}

\subsection{Randomness from single-qubit white noise and uncertainty relations}
\label{sec_uncertain}

Among the many possible illustrations of randomness in the tomographic level of characterization, we consider the case of randomness extraction from a single qubit in the maximally mixed state $\rho=\mathbb{1}/2$, against an adversary of Class I(a). We recover known results on uncertainty relations and show numerical evidence for more general situations.

As mentioned at the beginning of paragraph \ref{sec_tomo}, for a single measurement [Eq.~\eqref{eq_guessing_tomo_2}] one has $G(\mathbb{1}/2)=0$: for whatever measurement being performed being performed, the mixture may have been prepared by mixing the two eigenstates of the measurement and this information could be available to Eve. So to bound the guessing capability of Eve, we need to consider \textit{more than one measurements}. We thus refer to Eqs \eqref{eq_guessing_tomo_2gen} and \eqref{eq_guessing_tomo_3} from now onwards.

Let us denote $\{M_k\}_{k=1,...,N}$ the $N$ projective measurements: for any string of values $C=(c_1,...,c_N)\in\{-1,+1\}^N$, the effective measurement operator is
\ba
{\cal M}_C=\frac{\mathbb{1}+\vec{n}_C\cdot\vec{\sigma}}{2}&\textrm{  with  }& \vec{n}_C=\sum_{k=1}^N q_k\,c_k\,\vec{n}_k\ .
\ea
With this notation, we have
\begin{equation}
G(\mathbb{1}/2,\{M_k\})=\frac{1+\max_C|\vec{n}_C|}{2}\ .
\end{equation}
Indeed, the r.h.s. is obviously an upper bound, since it is the largest of the eigenvalues; and it can be achieved by the decomposition $\mathbb{1}/2=\frac{1}{2}\ket{+\vec{n}_{\bar{C}}}\bra{+\vec{n}_{\bar{C}}}+\frac{1}{2}\ket{-\vec{n}_{\bar{C}}}\bra{-\vec{n}_{\bar{C}}}$ where $\bar{C}$ is defined by $|\vec{n}_{\bar{C}}|=\max_C|\vec{n}_C|$. Finally, we are allowed to choose the $N$ most favorable measurements, i.e. Eq.~\eqref{eq_guessing_tomo_3} becomes $G(\mathbb{1}/2,N)=\frac{1+g_N}{2}$ for
\ba
g_N&\equiv&\min_{\{M_k,q_k\}} \max_C|\vec{n}_C|\,.
\ea
Now, since $|\vec{n}_k|=1$, we have
\ba |\vec{n}_C|^2=\sum_{k} q_k^2+\sum_{k\neq k'} (c_kq_k\vec{n}_k)\cdot(c_{k'}q_{k'}\vec{n}_{k'})\,.\ea Notice that the second term can always be made non-negative by the maximization over $C$. Indeed, it follows from
\ba
\sum_{c_1,...,c_N} \sum_{k\neq k'}(c_kq_k\vec{n}_k)\cdot(c_{k'}q_{k'}\vec{n}_{k'}) = 0
\ea
that $\max_{C} \sum_{k\neq k'}(c_kq_k\vec{n}_k)\cdot(c_{k'}q_{k'}\vec{n}_{k'})\geq0$. Therefore, in the minimization, the best choice would consist in choosing all the vectors mutually orthogonal, but this is possible only for $N=2,3$. In these cases, it is simple to finish the optimization: we find $g_2=1/\sqrt{2}\approx 0.7071$ and $g_3=1/\sqrt{3}\approx 0.5774$. Notice that, translated in min-entropy, the case for $N=2$ bound saturates the uncertainty relation for two min-entropies, Eq.~(9) of \cite{maassenuffink}, namely $H_{\min}(\sigma_z)+H_{\min}(\sigma_x)\geq\log(\frac{1+1/\sqrt{2}}{2})$.

To go further, we resort to numerical optimization to obtain upper bounds on $g_N$. For $N=4$, the optimal choice of measurements is found to be $\{(\sigma_z,1-3q), (\sigma_1,q), (\sigma_2,q), (-\sigma_3,q)\}$ where the vectors $\vec{n}_{1,2,3}$ are 120 degrees apart from each other in the $x-y$ plane; knowing this geometry, one can finish the optimization analytically to find $g_4=\sqrt{4/13}\approx 0.5547$ for $q=3/13$. For $N=5$ and $N=6$, we find $g_5\approx 0.5422$ and $g_6\approx 0.5270$. This trend suggests that, when $N\rightarrow\infty$, one has $g_N\rightarrow\frac{1}{2}$; we checked that this would be the case if the optimal choice would consist in spreading the $\vec{n}_k$ uniformly in the half-sphere.

\subsection{Randomness from POVMs}
\label{sec_povm}

As we have just seen and also mentioned in paragraph \ref{sec_tomo}, no randomness can be extracted from a single projective measurement on the maximally mixed state, because Eve may know the decomposition in the eigenvalues of that measurement. The reasoning does not seem to apply to POVMs, though: even knowing a pure state, in general Eve cannot guess with certainty the outcome of a non-projective POVM. Is it therefore possible to extract randomness from a single POVM on the maximally mixed state? The answer is, yes, but the origin of the randomness makes the problem trivial. Indeed, because of Neumark's theorem, a POVM is nothing else than a projective measurement on the system and some additional degrees of freedom. We are going to show that a POVM on the maximally mixed state cannot provide more randomness than that present in the ancilla --- thence it is pointless to perform the POVM for randomness purposes, one could have measured the ancilla directly.

To see this, let us first denote the system's dimensions by $d_s$. The n-outcomes POVM elemenets are written as $\{ \Pi_c \}_{c=1,...,n}$. Given that the state we have is white noise, the probability for Eve to guess the outcomes correctly is
\begin{eqnarray}
&&G(\mathbb{1}/d_s, \{ \Pi_c \})\nonumber\\	&=& \max_{ \{q_c, \ket{\psi_c}\}} \sum_{c=1}^n q_c \tr \big[ \rho_c \Pi_c \big] \label{eq_povm_summation}
\end{eqnarray}
where $\mathbb{1}/d_s = \sum_c q_c \rho_c$ and, as previously, $\rho_c$ groups all the states in the decomposition for which $\max_{c'}\tr [ \rho \Pi_{c'}]=\tr [ \rho \Pi_{c}]$. Since the maximization in Equation \eqref{eq_povm_summation} is taken over all possible decomposition, any specific decomposition provides a lower bound. In particular, we can choose
\begin{eqnarray}
	\rho_c = \frac{\Pi_c}{\tr \big[ \Pi_c \big] } &, \; \; & q_c = \frac{\tr \big[ \Pi_c \big] }{d_s}
\end{eqnarray}
which indeed gives $\sum_c q_c \rho_c = \mathbb{1}/d_s$. Then \begin{eqnarray}
	&&G(\mathbb{1}/d_s, \{ \Pi_c \}) \nonumber \\
	&\geq&  \sum_{c=1}^n \frac{ \tr \big[ \Pi_c \big] }{ d_s }  \tr \bigg[ \frac{ \Pi_c }{ \tr [ \Pi_c ] } \Pi_c \bigg] \nonumber \\
	&=& \sum_{c=1}^n \frac{ 1 }{ d_s }  \tr \big[  \Pi_c^2 \big]\,. \label{eq_povm_2}
\end{eqnarray}
Each $\Pi_c$ can be written as its spectral decomposition
\begin{eqnarray}
	\Pi_c &=& \sum_{k=1}^{r_c} \mu_{c,k} \ket{k_c} \bra{k_c}
\end{eqnarray}
where $r_c = \mathrm{rank} (\Pi_c)$ and $\mu_{c,k}$ are the eigenvalues of $\Pi_c$, with $\ket{k_c} \bra{k_c}$ to be the corresponding projector. Substituting the spectral decomposition into Eq.~\eqref{eq_povm_2} gives
\begin{eqnarray}
 \sum_{c=1}^n \frac{ 1 }{ d_s }  \tr \big[  \Pi_c^2 \big]	&=& \frac{ 1 }{ d_s }  \sum_{c=1}^n \sum_{k=1}^{r_c} \mu_{c,k}^2 \label{eq_povm_3} 
\end{eqnarray}
But using the Cauchy-Schwarz inequality, we have
\begin{eqnarray}
	\bigg[ \sum_{c=1}^n \sum_{k=1}^{r_c} \mu_{k,c}^2 \bigg] \bigg[ \sum_{c=1}^n \sum_{k=1}^{r_c} 1^2 \bigg]  &\geq& \bigg[ \sum_{c=1}^n \sum_{k=1}^{r_c} \mu_{c,k} \bigg]^2 \nonumber
\end{eqnarray} that is,
\begin{eqnarray}
	\sum_{c=1}^n \sum_{k=1}^{r_c} \mu_{c,k}^2  &\geq& \frac{d_s^2}{\sum_{c=1}^n r_c},
\end{eqnarray}
because $\sum_{c=1}^n \sum_{k=1}^{r_c} \mu_{c,k} = \sum_{c=1}^n \tr \big( \Pi_k \big) = d_s$. By substituting this inequality into Eq.~\eqref{eq_povm_3}, we find finally the following lower bound on the guessing probability:
\begin{eqnarray}
	G(\mathbb{1}/d_s, \{ \Pi_c \}) &\geq& \frac{d_s}{\sum_{c=1}^n r_c}
\end{eqnarray}
that is, the upper bound on the min-entropy
\ba
H_{\min}(\mathbb{1}/d_s, \{ \Pi_c \})&\leq& \log(\sum_c {r_c})-\log d_s\,.
\ea
In the tensor product implementation of the POVM, which uses an ancilla of dimension $d_a$, we have $\sum_c {r_c}=d_s\,d_a$; whence the maximum min-entropy is $\log d_a$, which comes solely from the ancilla. In the direct sum implementation, $\sum_c {r_c}=d_s+d_h$ is the minimum total dimension (system + hidden subspace) required to implement the POVM~\cite{chen07povm}: since $\log(d_s+d_h)\leq \log d_s+\log d_h$, the min-entropy is upper bounded by $\log d_h$. In both cases, we have proved our claim: all the randomness that can be obtained in a POVM on the maximally mixed state can be ascribed to the additional degrees of freedom used to implement the POVM. Finally notice that, as far as our proof goes, this conclusion applies only when the system is in the maximally mixed state: it remains an open problem whether, in other cases, POVMs may extract more randomness from the system than projective measurements. 

\subsection{Randomness from pointer measurements}

The previous observation on POVMs extends to another case in which additional degrees of freedom are used: that of measurement by coupling the relevant degree of freedom to a pointer. The best known textbook example is the Stern-Gerlach experiment. More common nowadays is the \textit{measurement of the polarization of a photon}: the photon is sent on a polarizing beam-splitter (PBS), the two output ports of which are correlated with orthogonal polarizations. It is by detecting in which beam the photon is (pointer) that polarization is inferred. Now, if one sets up an experiment to extract randomness from the polarization qubit \cite{fiorentino07,vallone}, the same setup provides another degree of freedom, whose state must be very close to pure if the measurement has to make sense at all: indeed, the beam must come from a well-defined direction for the PBS to work as expected. Thence, in principle one can extract more randomness by ignoring polarization and sending the photon on a normal beam-splitter \cite{jennewein00,stefanov00}. We stress that this argument bears on the amount of randomness and on simplicity ``on paper": polarization may be preferable to deal with other practical concerns \cite{fiorentino07}.

For other qubits, things may be more subtle. Consider for instance the probing of an atomic qubit with a laser beam: a laser beam alone can be used to generate randomness \cite{symul11}, but with a different detection scheme than the one used in probing atomic excitations; so it may not be immediate to suggest that one should ignore the atom and extract randomness directly from the laser. For yet other pointer measurements, it may not even be feasible to measure the pointer in a complementary basis (certainly it would be challenging for the Stern-Gerlach setup).

It is not our aim to propose concrete schemes to extract randomness at the tomography level of characterization. Rather, the bottom-line message could be put this way: whenever a quantum degree of freedom is measured by coupling it to a pointer, the pointer is usually in a well-defined quantum state. So, \textit{if the goal is to extract randomness, it is worth considering the possibility of getting it directly from the pointer}.

\section{Conclusion} In this work, we have quantified the randomness that can be extracted from a given quantum device with the device-independent, one-sided device-independent, and tomographic level of characterization. Specific tools were introduced to perform this quantification in the one-sided device-independent and tomographic cases. For the latter, we have also shown that not all conceivable procedures to extract randomness are actually relevant: in particular one must be careful whenever ancillas are involved since the randomness may just come from them rather than from the system under study. We have focused on the minimal class of adversarial power, relevant for the study of implementations performed by trusted experimentalists; a similar study could be conducted for randomness extraction against more powerful adversaries.  

\section*{Acknowledgment}

This work is funded by the Singapore Ministry of Education (partly through the Academic Research Fund Tier 3 MOE2012-T3-1-009) and by the Singapore National Research Foundation. We thank two anonymous referees for useful comments that lead to improving the structure of the paper, and Gonzalo de la Torre for discussions.

\bibliography{reference}

\begin{thebibliography}{61}%
\makeatletter
\providecommand \@ifxundefined [1]{%
 \@ifx{#1\undefined}
}%
\providecommand \@ifnum [1]{%
 \ifnum #1\expandafter \@firstoftwo
 \else \expandafter \@secondoftwo
 \fi
}%
\providecommand \@ifx [1]{%
 \ifx #1\expandafter \@firstoftwo
 \else \expandafter \@secondoftwo
 \fi
}%
\providecommand \natexlab [1]{#1}%
\providecommand \enquote  [1]{``#1''}%
\providecommand \bibnamefont  [1]{#1}%
\providecommand \bibfnamefont [1]{#1}%
\providecommand \citenamefont [1]{#1}%
\providecommand \href@noop [0]{\@secondoftwo}%
\providecommand \href [0]{\begingroup \@sanitize@url \@href}%
\providecommand \@href[1]{\@@startlink{#1}\@@href}%
\providecommand \@@href[1]{\endgroup#1\@@endlink}%
\providecommand \@sanitize@url [0]{\catcode `\\12\catcode `\$12\catcode
  `\&12\catcode `\#12\catcode `\^12\catcode `\_12\catcode `\%12\relax}%
\providecommand \@@startlink[1]{}%
\providecommand \@@endlink[0]{}%
\providecommand \url  [0]{\begingroup\@sanitize@url \@url }%
\providecommand \@url [1]{\endgroup\@href {#1}{\urlprefix }}%
\providecommand \urlprefix  [0]{URL }%
\providecommand \Eprint [0]{\href }%
\providecommand \doibase [0]{http://dx.doi.org/}%
\providecommand \selectlanguage [0]{\@gobble}%
\providecommand \bibinfo  [0]{\@secondoftwo}%
\providecommand \bibfield  [0]{\@secondoftwo}%
\providecommand \translation [1]{[#1]}%
\providecommand \BibitemOpen [0]{}%
\providecommand \bibitemStop [0]{}%
\providecommand \bibitemNoStop [0]{.\EOS\space}%
\providecommand \EOS [0]{\spacefactor3000\relax}%
\providecommand \BibitemShut  [1]{\csname bibitem#1\endcsname}%
\let\auto@bib@innerbib\@empty
\bibitem [{\citenamefont {Bell}(1964)}]{bell1964}%
  \BibitemOpen
  \bibfield  {author} {\bibinfo {author} {\bibfnamefont {J.~S.}\ \bibnamefont
  {Bell}},\ }\href@noop {} {\bibfield  {journal} {\bibinfo  {journal} {Physics
  (Long Island City, N.Y.)}\ }\textbf {\bibinfo {volume} {1}},\ \bibinfo
  {pages} {195} (\bibinfo {year} {1964})}\BibitemShut {NoStop}%
\bibitem [{\citenamefont {Scarani}\ and\ \citenamefont {Gisin}(2002)}]{SG02}%
  \BibitemOpen
  \bibfield  {author} {\bibinfo {author} {\bibfnamefont {V.}~\bibnamefont
  {Scarani}}\ and\ \bibinfo {author} {\bibfnamefont {N.}~\bibnamefont
  {Gisin}},\ }\href@noop {} {\bibfield  {journal} {\bibinfo  {journal}
  {Physical Review A}\ }\textbf {\bibinfo {volume} {65}},\ \bibinfo {pages}
  {012311} (\bibinfo {year} {2002})}\BibitemShut {NoStop}%
\bibitem [{\citenamefont {Brukner}\ \emph {et~al.}(2004)\citenamefont
  {Brukner}, \citenamefont {Zukowski}, \citenamefont {Pan},\ and\ \citenamefont
  {Zeilinger}}]{vienna04}%
  \BibitemOpen
  \bibfield  {author} {\bibinfo {author} {\bibfnamefont {C.}~\bibnamefont
  {Brukner}}, \bibinfo {author} {\bibfnamefont {M.}~\bibnamefont {Zukowski}},
  \bibinfo {author} {\bibfnamefont {J.~W.}\ \bibnamefont {Pan}}, \ and\
  \bibinfo {author} {\bibfnamefont {A.}~\bibnamefont {Zeilinger}},\ }\href@noop
  {} {\bibfield  {journal} {\bibinfo  {journal} {Physical Review Letters}\
  }\textbf {\bibinfo {volume} {92}},\ \bibinfo {pages} {127901} (\bibinfo
  {year} {2004})}\BibitemShut {NoStop}%
\bibitem [{\citenamefont {Brunner}\ \emph {et~al.}(2014)\citenamefont
  {Brunner}, \citenamefont {Cavalcanti}, \citenamefont {Pironio}, \citenamefont
  {Scarani},\ and\ \citenamefont {Wehner}}]{review}%
  \BibitemOpen
  \bibfield  {author} {\bibinfo {author} {\bibfnamefont {N.}~\bibnamefont
  {Brunner}}, \bibinfo {author} {\bibfnamefont {D.}~\bibnamefont {Cavalcanti}},
  \bibinfo {author} {\bibfnamefont {S.}~\bibnamefont {Pironio}}, \bibinfo
  {author} {\bibfnamefont {V.}~\bibnamefont {Scarani}}, \ and\ \bibinfo
  {author} {\bibfnamefont {S.}~\bibnamefont {Wehner}},\ }\href@noop {}
  {\bibfield  {journal} {\bibinfo  {journal} {Reviews of Modern Physics}\
  }\textbf {\bibinfo {volume} {86}},\ \bibinfo {pages} {012311} (\bibinfo
  {year} {2014})}\BibitemShut {NoStop}%
\bibitem [{\citenamefont {Scarani}(2012)}]{slovaca}%
  \BibitemOpen
  \bibfield  {author} {\bibinfo {author} {\bibfnamefont {V.}~\bibnamefont
  {Scarani}},\ }\href@noop {} {\bibfield  {journal} {\bibinfo  {journal} {Acta
  Physica Slovaca}\ }\textbf {\bibinfo {volume} {62}},\ \bibinfo {pages} {247}
  (\bibinfo {year} {2012})}\BibitemShut {NoStop}%
\bibitem [{\citenamefont {Acin}\ \emph {et~al.}(2007)\citenamefont {Acin},
  \citenamefont {Brunner}, \citenamefont {Gisin}, \citenamefont {Massar},
  \citenamefont {Pironio},\ and\ \citenamefont {Scarani}}]{acin2007device}%
  \BibitemOpen
  \bibfield  {author} {\bibinfo {author} {\bibfnamefont {A.}~\bibnamefont
  {Acin}}, \bibinfo {author} {\bibfnamefont {N.}~\bibnamefont {Brunner}},
  \bibinfo {author} {\bibfnamefont {N.}~\bibnamefont {Gisin}}, \bibinfo
  {author} {\bibfnamefont {S.}~\bibnamefont {Massar}}, \bibinfo {author}
  {\bibfnamefont {S.}~\bibnamefont {Pironio}}, \ and\ \bibinfo {author}
  {\bibfnamefont {V.}~\bibnamefont {Scarani}},\ }\href@noop {} {\bibfield
  {journal} {\bibinfo  {journal} {Physical Review Letters}\ }\textbf {\bibinfo
  {volume} {98}},\ \bibinfo {pages} {230501} (\bibinfo {year}
  {2007})}\BibitemShut {NoStop}%
\bibitem [{\citenamefont {Pironio}\ \emph {et~al.}(2010)\citenamefont
  {Pironio}, \citenamefont {Ac{\'\i}n}, \citenamefont {Massar}, \citenamefont
  {de~La~Giroday}, \citenamefont {Matsukevich}, \citenamefont {Maunz},
  \citenamefont {Olmschenk}, \citenamefont {Hayes}, \citenamefont {Luo},
  \citenamefont {Manning} \emph {et~al.}}]{pironio10random}%
  \BibitemOpen
  \bibfield  {author} {\bibinfo {author} {\bibfnamefont {S.}~\bibnamefont
  {Pironio}}, \bibinfo {author} {\bibfnamefont {A.}~\bibnamefont {Ac{\'\i}n}},
  \bibinfo {author} {\bibfnamefont {S.}~\bibnamefont {Massar}}, \bibinfo
  {author} {\bibfnamefont {A.~B.}\ \bibnamefont {de~La~Giroday}}, \bibinfo
  {author} {\bibfnamefont {D.~N.}\ \bibnamefont {Matsukevich}}, \bibinfo
  {author} {\bibfnamefont {P.}~\bibnamefont {Maunz}}, \bibinfo {author}
  {\bibfnamefont {S.}~\bibnamefont {Olmschenk}}, \bibinfo {author}
  {\bibfnamefont {D.}~\bibnamefont {Hayes}}, \bibinfo {author} {\bibfnamefont
  {L.}~\bibnamefont {Luo}}, \bibinfo {author} {\bibfnamefont {T.~A.}\
  \bibnamefont {Manning}},  \emph {et~al.},\ }\href@noop {} {\bibfield
  {journal} {\bibinfo  {journal} {Nature}\ }\textbf {\bibinfo {volume} {464}},\
  \bibinfo {pages} {1021} (\bibinfo {year} {2010})}\BibitemShut {NoStop}%
\bibitem [{\citenamefont {Colbeck}\ and\ \citenamefont
  {Kent}(2011)}]{colbeck11private}%
  \BibitemOpen
  \bibfield  {author} {\bibinfo {author} {\bibfnamefont {R.}~\bibnamefont
  {Colbeck}}\ and\ \bibinfo {author} {\bibfnamefont {A.}~\bibnamefont {Kent}},\
  }\href@noop {} {\bibfield  {journal} {\bibinfo  {journal} {Journal of Physics
  A: Mathematical and Theoretical}\ }\textbf {\bibinfo {volume} {44}},\
  \bibinfo {pages} {095305} (\bibinfo {year} {2011})}\BibitemShut {NoStop}%
\bibitem [{\citenamefont {Bardyn}\ \emph {et~al.}(2009)\citenamefont {Bardyn},
  \citenamefont {Liew}, \citenamefont {Massar}, \citenamefont {McKague},\ and\
  \citenamefont {Scarani}}]{bardyn09}%
  \BibitemOpen
  \bibfield  {author} {\bibinfo {author} {\bibfnamefont {C.-E.}\ \bibnamefont
  {Bardyn}}, \bibinfo {author} {\bibfnamefont {T.~C.}\ \bibnamefont {Liew}},
  \bibinfo {author} {\bibfnamefont {S.}~\bibnamefont {Massar}}, \bibinfo
  {author} {\bibfnamefont {M.}~\bibnamefont {McKague}}, \ and\ \bibinfo
  {author} {\bibfnamefont {V.}~\bibnamefont {Scarani}},\ }\href
  {http://journals.aps.org/pra/abstract/10.1103/PhysRevA.80.062327} {\bibfield
  {journal} {\bibinfo  {journal} {Phys. Rev. A}\ }\textbf {\bibinfo {volume}
  {80}},\ \bibinfo {pages} {062327} (\bibinfo {year} {2009})}\BibitemShut
  {NoStop}%
\bibitem [{\citenamefont {Moroder}\ \emph {et~al.}(2013)\citenamefont
  {Moroder}, \citenamefont {Bancal}, \citenamefont {Liang}, \citenamefont
  {Hofmann},\ and\ \citenamefont {G\"uhne}}]{moroder2013}%
  \BibitemOpen
  \bibfield  {author} {\bibinfo {author} {\bibfnamefont {T.}~\bibnamefont
  {Moroder}}, \bibinfo {author} {\bibfnamefont {J.-D.}\ \bibnamefont {Bancal}},
  \bibinfo {author} {\bibfnamefont {Y.-C.}\ \bibnamefont {Liang}}, \bibinfo
  {author} {\bibfnamefont {M.}~\bibnamefont {Hofmann}}, \ and\ \bibinfo
  {author} {\bibfnamefont {O.}~\bibnamefont {G\"uhne}},\ }\href {\doibase
  10.1103/PhysRevLett.111.030501} {\bibfield  {journal} {\bibinfo  {journal}
  {Phys. Rev. Lett.}\ }\textbf {\bibinfo {volume} {111}},\ \bibinfo {pages}
  {030501} (\bibinfo {year} {2013})}\BibitemShut {NoStop}%
\bibitem [{\citenamefont {Ac\'{\i}n}\ \emph {et~al.}(2006)\citenamefont
  {Ac\'{\i}n}, \citenamefont {Gisin},\ and\ \citenamefont {Masanes}}]{agm06}%
  \BibitemOpen
  \bibfield  {author} {\bibinfo {author} {\bibfnamefont {A.}~\bibnamefont
  {Ac\'{\i}n}}, \bibinfo {author} {\bibfnamefont {N.}~\bibnamefont {Gisin}}, \
  and\ \bibinfo {author} {\bibfnamefont {L.}~\bibnamefont {Masanes}},\
  }\href@noop {} {\bibfield  {journal} {\bibinfo  {journal} {Physical Review
  Letters}\ }\textbf {\bibinfo {volume} {97}},\ \bibinfo {pages} {120405}
  (\bibinfo {year} {2006})}\BibitemShut {NoStop}%
\bibitem [{\citenamefont {Colbeck}\ and\ \citenamefont
  {Renner}(2012)}]{colbeck12amplifysv}%
  \BibitemOpen
  \bibfield  {author} {\bibinfo {author} {\bibfnamefont {R.}~\bibnamefont
  {Colbeck}}\ and\ \bibinfo {author} {\bibfnamefont {R.}~\bibnamefont
  {Renner}},\ }\href@noop {} {\bibfield  {journal} {\bibinfo  {journal} {Nature
  Physics}\ }\textbf {\bibinfo {volume} {8}},\ \bibinfo {pages} {450} (\bibinfo
  {year} {2012})}\BibitemShut {NoStop}%
\bibitem [{\citenamefont {{Grudka}}\ \emph {et~al.}(2013)\citenamefont
  {{Grudka}}, \citenamefont {{Horodecki}}, \citenamefont {{Horodecki}},
  \citenamefont {{Horodecki}}, \citenamefont {{Paw{\l}owski}},\ and\
  \citenamefont {{Ramanathan}}}]{grudka13amplifysv}%
  \BibitemOpen
  \bibfield  {author} {\bibinfo {author} {\bibfnamefont {A.}~\bibnamefont
  {{Grudka}}}, \bibinfo {author} {\bibfnamefont {K.}~\bibnamefont
  {{Horodecki}}}, \bibinfo {author} {\bibfnamefont {M.}~\bibnamefont
  {{Horodecki}}}, \bibinfo {author} {\bibfnamefont {P.}~\bibnamefont
  {{Horodecki}}}, \bibinfo {author} {\bibfnamefont {M.}~\bibnamefont
  {{Paw{\l}owski}}}, \ and\ \bibinfo {author} {\bibfnamefont {R.}~\bibnamefont
  {{Ramanathan}}},\ }\href@noop {} {\bibfield  {journal} {\bibinfo  {journal}
  {ArXiv e-prints}\ } (\bibinfo {year} {2013})},\ \Eprint
  {http://arxiv.org/abs/1303.5591} {arXiv:1303.5591 [quant-ph]} \BibitemShut
  {NoStop}%
\bibitem [{\citenamefont {Gallego}\ \emph {et~al.}(2013)\citenamefont
  {Gallego}, \citenamefont {Masanes}, \citenamefont {De~La~Torre},
  \citenamefont {Dhara}, \citenamefont {Aolita},\ and\ \citenamefont
  {Ac{\'\i}n}}]{gallego13amplifysv}%
  \BibitemOpen
  \bibfield  {author} {\bibinfo {author} {\bibfnamefont {R.}~\bibnamefont
  {Gallego}}, \bibinfo {author} {\bibfnamefont {L.}~\bibnamefont {Masanes}},
  \bibinfo {author} {\bibfnamefont {G.}~\bibnamefont {De~La~Torre}}, \bibinfo
  {author} {\bibfnamefont {C.}~\bibnamefont {Dhara}}, \bibinfo {author}
  {\bibfnamefont {L.}~\bibnamefont {Aolita}}, \ and\ \bibinfo {author}
  {\bibfnamefont {A.}~\bibnamefont {Ac{\'\i}n}},\ }\href
  {http://dx.doi.org/10.1038/ncomms3654} {\bibfield  {journal} {\bibinfo
  {journal} {Nat Commun}\ }\textbf {\bibinfo {volume} {4}} (\bibinfo {year}
  {2013})},\ \bibinfo {note} {article}\BibitemShut {NoStop}%
\bibitem [{\citenamefont {{Ramanathan}}\ \emph {et~al.}(2013)\citenamefont
  {{Ramanathan}}, \citenamefont {{Brandao}}, \citenamefont {{Grudka}},
  \citenamefont {{Horodecki}}, \citenamefont {{Horodecki}},\ and\ \citenamefont
  {{Horodecki}}}]{ramanathan13amplifysv}%
  \BibitemOpen
  \bibfield  {author} {\bibinfo {author} {\bibfnamefont {R.}~\bibnamefont
  {{Ramanathan}}}, \bibinfo {author} {\bibfnamefont {F.~G.~S.~L.}\ \bibnamefont
  {{Brandao}}}, \bibinfo {author} {\bibfnamefont {A.}~\bibnamefont {{Grudka}}},
  \bibinfo {author} {\bibfnamefont {K.}~\bibnamefont {{Horodecki}}}, \bibinfo
  {author} {\bibfnamefont {M.}~\bibnamefont {{Horodecki}}}, \ and\ \bibinfo
  {author} {\bibfnamefont {P.}~\bibnamefont {{Horodecki}}},\ }\href@noop {}
  {\bibfield  {journal} {\bibinfo  {journal} {ArXiv e-prints}\ } (\bibinfo
  {year} {2013})},\ \Eprint {http://arxiv.org/abs/1308.4635} {arXiv:1308.4635
  [quant-ph]} \BibitemShut {NoStop}%
\bibitem [{\citenamefont {{Brand{\~a}o}}\ \emph {et~al.}(2013)\citenamefont
  {{Brand{\~a}o}}, \citenamefont {{Ramanathan}}, \citenamefont {{Grudka}},
  \citenamefont {{Horodecki}}, \citenamefont {{Horodecki}},\ and\ \citenamefont
  {{Horodecki}}}]{brandao13amplifysv}%
  \BibitemOpen
  \bibfield  {author} {\bibinfo {author} {\bibfnamefont {F.~G.~S.~L.}\
  \bibnamefont {{Brand{\~a}o}}}, \bibinfo {author} {\bibfnamefont
  {R.}~\bibnamefont {{Ramanathan}}}, \bibinfo {author} {\bibfnamefont
  {A.}~\bibnamefont {{Grudka}}}, \bibinfo {author} {\bibfnamefont
  {K.}~\bibnamefont {{Horodecki}}}, \bibinfo {author} {\bibfnamefont
  {M.}~\bibnamefont {{Horodecki}}}, \ and\ \bibinfo {author} {\bibfnamefont
  {P.}~\bibnamefont {{Horodecki}}},\ }\href@noop {} {\bibfield  {journal}
  {\bibinfo  {journal} {ArXiv e-prints}\ } (\bibinfo {year} {2013})},\ \Eprint
  {http://arxiv.org/abs/1310.4544} {arXiv:1310.4544 [quant-ph]} \BibitemShut
  {NoStop}%
\bibitem [{\citenamefont {{Mironowicz}}\ \emph {et~al.}(2013)\citenamefont
  {{Mironowicz}}, \citenamefont {{Gallego}},\ and\ \citenamefont
  {{Pawlowski}}}]{mironowicz13amplifysv}%
  \BibitemOpen
  \bibfield  {author} {\bibinfo {author} {\bibfnamefont {P.}~\bibnamefont
  {{Mironowicz}}}, \bibinfo {author} {\bibfnamefont {R.}~\bibnamefont
  {{Gallego}}}, \ and\ \bibinfo {author} {\bibfnamefont {M.}~\bibnamefont
  {{Pawlowski}}},\ }\href@noop {} {\bibfield  {journal} {\bibinfo  {journal}
  {ArXiv e-prints}\ } (\bibinfo {year} {2013})},\ \Eprint
  {http://arxiv.org/abs/1301.7722} {arXiv:1301.7722 [quant-ph]} \BibitemShut
  {NoStop}%
\bibitem [{\citenamefont {{Bouda}}\ \emph {et~al.}(2014)\citenamefont
  {{Bouda}}, \citenamefont {{Pawlowski}}, \citenamefont {{Pivoluska}},\ and\
  \citenamefont {{Plesch}}}]{bouda14amplifymin}%
  \BibitemOpen
  \bibfield  {author} {\bibinfo {author} {\bibfnamefont {J.}~\bibnamefont
  {{Bouda}}}, \bibinfo {author} {\bibfnamefont {M.}~\bibnamefont
  {{Pawlowski}}}, \bibinfo {author} {\bibfnamefont {M.}~\bibnamefont
  {{Pivoluska}}}, \ and\ \bibinfo {author} {\bibfnamefont {M.}~\bibnamefont
  {{Plesch}}},\ }\href@noop {} {\bibfield  {journal} {\bibinfo  {journal}
  {ArXiv e-prints}\ } (\bibinfo {year} {2014})},\ \Eprint
  {http://arxiv.org/abs/1402.0974} {arXiv:1402.0974 [quant-ph]} \BibitemShut
  {NoStop}%
\bibitem [{\citenamefont {{Chung}}\ \emph {et~al.}(2014)\citenamefont
  {{Chung}}, \citenamefont {{Shi}},\ and\ \citenamefont
  {{Wu}}}]{chung14amplifymin}%
  \BibitemOpen
  \bibfield  {author} {\bibinfo {author} {\bibfnamefont {K.-M.}\ \bibnamefont
  {{Chung}}}, \bibinfo {author} {\bibfnamefont {Y.}~\bibnamefont {{Shi}}}, \
  and\ \bibinfo {author} {\bibfnamefont {X.}~\bibnamefont {{Wu}}},\ }\href@noop
  {} {\bibfield  {journal} {\bibinfo  {journal} {ArXiv e-prints}\ } (\bibinfo
  {year} {2014})},\ \Eprint {http://arxiv.org/abs/1402.4797} {arXiv:1402.4797
  [quant-ph]} \BibitemShut {NoStop}%
\bibitem [{\citenamefont {Ac{\'\i}n}\ \emph {et~al.}(2006)\citenamefont
  {Ac{\'\i}n}, \citenamefont {Gisin},\ and\ \citenamefont {Masanes}}]{acin06}%
  \BibitemOpen
  \bibfield  {author} {\bibinfo {author} {\bibfnamefont {A.}~\bibnamefont
  {Ac{\'\i}n}}, \bibinfo {author} {\bibfnamefont {N.}~\bibnamefont {Gisin}}, \
  and\ \bibinfo {author} {\bibfnamefont {L.}~\bibnamefont {Masanes}},\ }\href
  {\doibase 10.1103/PhysRevLett.97.120405} {\bibfield  {journal} {\bibinfo
  {journal} {Phys. Rev. Lett.}\ }\textbf {\bibinfo {volume} {97}},\ \bibinfo
  {pages} {120405} (\bibinfo {year} {2006})}\BibitemShut {NoStop}%
\bibitem [{\citenamefont {{Coudron}}\ and\ \citenamefont
  {{Yuen}}(2013)}]{coudron13expand}%
  \BibitemOpen
  \bibfield  {author} {\bibinfo {author} {\bibfnamefont {M.}~\bibnamefont
  {{Coudron}}}\ and\ \bibinfo {author} {\bibfnamefont {H.}~\bibnamefont
  {{Yuen}}},\ }\href@noop {} {\bibfield  {journal} {\bibinfo  {journal} {ArXiv
  e-prints}\ } (\bibinfo {year} {2013})},\ \Eprint
  {http://arxiv.org/abs/1310.6755} {arXiv:1310.6755 [quant-ph]} \BibitemShut
  {NoStop}%
\bibitem [{\citenamefont {Damg{\aa}rd}\ \emph {et~al.}(2008)\citenamefont
  {Damg{\aa}rd}, \citenamefont {Fehr}, \citenamefont {Salvail},\ and\
  \citenamefont {Schaffner}}]{damgard08}%
  \BibitemOpen
  \bibfield  {author} {\bibinfo {author} {\bibfnamefont {I.~B.}\ \bibnamefont
  {Damg{\aa}rd}}, \bibinfo {author} {\bibfnamefont {S.}~\bibnamefont {Fehr}},
  \bibinfo {author} {\bibfnamefont {L.}~\bibnamefont {Salvail}}, \ and\
  \bibinfo {author} {\bibfnamefont {C.}~\bibnamefont {Schaffner}},\ }\href@noop
  {} {\bibfield  {journal} {\bibinfo  {journal} {SIAM Journal on Computing}\
  }\textbf {\bibinfo {volume} {37}},\ \bibinfo {pages} {1865} (\bibinfo {year}
  {2008})}\BibitemShut {NoStop}%
\bibitem [{\citenamefont {Schaffner}\ \emph {et~al.}(2009)\citenamefont
  {Schaffner}, \citenamefont {Terhal},\ and\ \citenamefont
  {Wehner}}]{wehner08noisy}%
  \BibitemOpen
  \bibfield  {author} {\bibinfo {author} {\bibfnamefont {C.}~\bibnamefont
  {Schaffner}}, \bibinfo {author} {\bibfnamefont {B.}~\bibnamefont {Terhal}}, \
  and\ \bibinfo {author} {\bibfnamefont {S.}~\bibnamefont {Wehner}},\ }\href
  {http://dl.acm.org/citation.cfm?id=2012098.2012102} {\bibfield  {journal}
  {\bibinfo  {journal} {Quantum Info. Comput.}\ }\textbf {\bibinfo {volume}
  {9}},\ \bibinfo {pages} {963} (\bibinfo {year} {2009})}\BibitemShut {NoStop}%
\bibitem [{\citenamefont {{Vazirani}}\ and\ \citenamefont
  {{Vidick}}(2011)}]{vazirani11expand}%
  \BibitemOpen
  \bibfield  {author} {\bibinfo {author} {\bibfnamefont {U.~V.}\ \bibnamefont
  {{Vazirani}}}\ and\ \bibinfo {author} {\bibfnamefont {T.}~\bibnamefont
  {{Vidick}}},\ }\href@noop {} {\bibfield  {journal} {\bibinfo  {journal}
  {ArXiv e-prints}\ } (\bibinfo {year} {2011})},\ \Eprint
  {http://arxiv.org/abs/1111.6054} {arXiv:1111.6054 [quant-ph]} \BibitemShut
  {NoStop}%
\bibitem [{\citenamefont {{Coudron}}\ \emph {et~al.}(2013)\citenamefont
  {{Coudron}}, \citenamefont {{Vidick}},\ and\ \citenamefont
  {{Yuen}}}]{coudron13}%
  \BibitemOpen
  \bibfield  {author} {\bibinfo {author} {\bibfnamefont {M.}~\bibnamefont
  {{Coudron}}}, \bibinfo {author} {\bibfnamefont {T.}~\bibnamefont {{Vidick}}},
  \ and\ \bibinfo {author} {\bibfnamefont {H.}~\bibnamefont {{Yuen}}},\
  }\href@noop {} {\bibfield  {journal} {\bibinfo  {journal} {ArXiv e-prints}\ }
  (\bibinfo {year} {2013})},\ \Eprint {http://arxiv.org/abs/1305.6626}
  {arXiv:1305.6626 [quant-ph]} \BibitemShut {NoStop}%
\bibitem [{\citenamefont {{Miller}}\ and\ \citenamefont
  {{Shi}}(2014)}]{miller14expand}%
  \BibitemOpen
  \bibfield  {author} {\bibinfo {author} {\bibfnamefont {C.~A.}\ \bibnamefont
  {{Miller}}}\ and\ \bibinfo {author} {\bibfnamefont {Y.}~\bibnamefont
  {{Shi}}},\ }\href@noop {} {\bibfield  {journal} {\bibinfo  {journal} {ArXiv
  e-prints}\ } (\bibinfo {year} {2014})},\ \Eprint
  {http://arxiv.org/abs/1402.0489} {arXiv:1402.0489 [quant-ph]} \BibitemShut
  {NoStop}%
\bibitem [{\citenamefont {Barrett}\ \emph {et~al.}(2013)\citenamefont
  {Barrett}, \citenamefont {Colbeck},\ and\ \citenamefont
  {Kent}}]{barrett13attackdiqkd}%
  \BibitemOpen
  \bibfield  {author} {\bibinfo {author} {\bibfnamefont {J.}~\bibnamefont
  {Barrett}}, \bibinfo {author} {\bibfnamefont {R.}~\bibnamefont {Colbeck}}, \
  and\ \bibinfo {author} {\bibfnamefont {A.}~\bibnamefont {Kent}},\ }\href
  {\doibase 10.1103/PhysRevLett.110.010503} {\bibfield  {journal} {\bibinfo
  {journal} {Phys. Rev. Lett.}\ }\textbf {\bibinfo {volume} {110}},\ \bibinfo
  {pages} {010503} (\bibinfo {year} {2013})}\BibitemShut {NoStop}%
\bibitem [{\citenamefont {Pironio}\ and\ \citenamefont
  {Massar}(2013)}]{pironio13expand}%
  \BibitemOpen
  \bibfield  {author} {\bibinfo {author} {\bibfnamefont {S.}~\bibnamefont
  {Pironio}}\ and\ \bibinfo {author} {\bibfnamefont {S.}~\bibnamefont
  {Massar}},\ }\href {\doibase 10.1103/PhysRevA.87.012336} {\bibfield
  {journal} {\bibinfo  {journal} {Phys. Rev. A}\ }\textbf {\bibinfo {volume}
  {87}},\ \bibinfo {pages} {012336} (\bibinfo {year} {2013})}\BibitemShut
  {NoStop}%
\bibitem [{\citenamefont {Fehr}\ \emph {et~al.}(2013)\citenamefont {Fehr},
  \citenamefont {Gelles},\ and\ \citenamefont {Schaffner}}]{fehr13expand}%
  \BibitemOpen
  \bibfield  {author} {\bibinfo {author} {\bibfnamefont {S.}~\bibnamefont
  {Fehr}}, \bibinfo {author} {\bibfnamefont {R.}~\bibnamefont {Gelles}}, \ and\
  \bibinfo {author} {\bibfnamefont {C.}~\bibnamefont {Schaffner}},\ }\href
  {\doibase 10.1103/PhysRevA.87.012335} {\bibfield  {journal} {\bibinfo
  {journal} {Phys. Rev. A}\ }\textbf {\bibinfo {volume} {87}},\ \bibinfo
  {pages} {012335} (\bibinfo {year} {2013})}\BibitemShut {NoStop}%
\bibitem [{\citenamefont {James}\ \emph {et~al.}(2001)\citenamefont {James},
  \citenamefont {Kwiat}, \citenamefont {Munro},\ and\ \citenamefont
  {White}}]{james01}%
  \BibitemOpen
  \bibfield  {author} {\bibinfo {author} {\bibfnamefont {D.~F.~V.}\
  \bibnamefont {James}}, \bibinfo {author} {\bibfnamefont {P.~G.}\ \bibnamefont
  {Kwiat}}, \bibinfo {author} {\bibfnamefont {W.~J.}\ \bibnamefont {Munro}}, \
  and\ \bibinfo {author} {\bibfnamefont {A.~G.}\ \bibnamefont {White}},\ }\href
  {\doibase 10.1103/PhysRevA.64.052312} {\bibfield  {journal} {\bibinfo
  {journal} {Phys. Rev. A}\ }\textbf {\bibinfo {volume} {64}},\ \bibinfo
  {pages} {052312} (\bibinfo {year} {2001})}\BibitemShut {NoStop}%
\bibitem [{\citenamefont {Moroder}\ \emph {et~al.}(2010)\citenamefont
  {Moroder}, \citenamefont {G\"uhne}, \citenamefont {Beaudry}, \citenamefont
  {Piani},\ and\ \citenamefont {L\"utkenhaus}}]{moroder10}%
  \BibitemOpen
  \bibfield  {author} {\bibinfo {author} {\bibfnamefont {T.}~\bibnamefont
  {Moroder}}, \bibinfo {author} {\bibfnamefont {O.}~\bibnamefont {G\"uhne}},
  \bibinfo {author} {\bibfnamefont {N.}~\bibnamefont {Beaudry}}, \bibinfo
  {author} {\bibfnamefont {M.}~\bibnamefont {Piani}}, \ and\ \bibinfo {author}
  {\bibfnamefont {N.}~\bibnamefont {L\"utkenhaus}},\ }\href {\doibase
  10.1103/PhysRevA.81.052342} {\bibfield  {journal} {\bibinfo  {journal} {Phys.
  Rev. A}\ }\textbf {\bibinfo {volume} {81}},\ \bibinfo {pages} {052342}
  (\bibinfo {year} {2010})}\BibitemShut {NoStop}%
\bibitem [{\citenamefont {Teo}\ \emph {et~al.}(2012)\citenamefont {Teo},
  \citenamefont {Stoklasa}, \citenamefont {Englert}, \citenamefont {\ifmmode
  \check{R}\else \v{R}\fi{}eh\'a\ifmmode~\check{c}\else \v{c}\fi{}ek},\ and\
  \citenamefont {Hradil}}]{teo12}%
  \BibitemOpen
  \bibfield  {author} {\bibinfo {author} {\bibfnamefont {Y.~S.}\ \bibnamefont
  {Teo}}, \bibinfo {author} {\bibfnamefont {B.}~\bibnamefont {Stoklasa}},
  \bibinfo {author} {\bibfnamefont {B.-G.}\ \bibnamefont {Englert}}, \bibinfo
  {author} {\bibfnamefont {J.}~\bibnamefont {\ifmmode \check{R}\else
  \v{R}\fi{}eh\'a\ifmmode~\check{c}\else \v{c}\fi{}ek}}, \ and\ \bibinfo
  {author} {\bibfnamefont {Z.~c.~v.}\ \bibnamefont {Hradil}},\ }\href {\doibase
  10.1103/PhysRevA.85.042317} {\bibfield  {journal} {\bibinfo  {journal} {Phys.
  Rev. A}\ }\textbf {\bibinfo {volume} {85}},\ \bibinfo {pages} {042317}
  (\bibinfo {year} {2012})}\BibitemShut {NoStop}%
\bibitem [{\citenamefont {Rosset}\ \emph {et~al.}(2012)\citenamefont {Rosset},
  \citenamefont {Ferretti-Sch\"obitz}, \citenamefont {Bancal}, \citenamefont
  {Gisin},\ and\ \citenamefont {Liang}}]{rosset12}%
  \BibitemOpen
  \bibfield  {author} {\bibinfo {author} {\bibfnamefont {D.}~\bibnamefont
  {Rosset}}, \bibinfo {author} {\bibfnamefont {R.}~\bibnamefont
  {Ferretti-Sch\"obitz}}, \bibinfo {author} {\bibfnamefont {J.-D.}\
  \bibnamefont {Bancal}}, \bibinfo {author} {\bibfnamefont {N.}~\bibnamefont
  {Gisin}}, \ and\ \bibinfo {author} {\bibfnamefont {Y.-C.}\ \bibnamefont
  {Liang}},\ }\href {\doibase 10.1103/PhysRevA.86.062325} {\bibfield  {journal}
  {\bibinfo  {journal} {Phys. Rev. A}\ }\textbf {\bibinfo {volume} {86}},\
  \bibinfo {pages} {062325} (\bibinfo {year} {2012})}\BibitemShut {NoStop}%
\bibitem [{\citenamefont {Lydersen}\ \emph {et~al.}(2010)\citenamefont
  {Lydersen}, \citenamefont {Wiechers}, \citenamefont {Wittmann}, \citenamefont
  {Elser}, \citenamefont {Skaar},\ and\ \citenamefont
  {Makarov}}]{lydersen10qkdhack}%
  \BibitemOpen
  \bibfield  {author} {\bibinfo {author} {\bibfnamefont {L.}~\bibnamefont
  {Lydersen}}, \bibinfo {author} {\bibfnamefont {C.}~\bibnamefont {Wiechers}},
  \bibinfo {author} {\bibfnamefont {C.}~\bibnamefont {Wittmann}}, \bibinfo
  {author} {\bibfnamefont {D.}~\bibnamefont {Elser}}, \bibinfo {author}
  {\bibfnamefont {J.}~\bibnamefont {Skaar}}, \ and\ \bibinfo {author}
  {\bibfnamefont {V.}~\bibnamefont {Makarov}},\ }\href {\doibase
  10.1038/nphoton.2010.214} {\bibfield  {journal} {\bibinfo  {journal} {Nat
  Photon}\ }\textbf {\bibinfo {volume} {4}},\ \bibinfo {pages} {686} (\bibinfo
  {year} {2010})}\BibitemShut {NoStop}%
\bibitem [{\citenamefont {Liu}\ \emph {et~al.}(2014)\citenamefont {Liu},
  \citenamefont {Lamas-Linares}, \citenamefont {Kurtsiefer}, \citenamefont
  {Skaar}, \citenamefont {Makarov},\ and\ \citenamefont
  {Gerhardt}}]{liu14qkdhack}%
  \BibitemOpen
  \bibfield  {author} {\bibinfo {author} {\bibfnamefont {Q.}~\bibnamefont
  {Liu}}, \bibinfo {author} {\bibfnamefont {A.}~\bibnamefont {Lamas-Linares}},
  \bibinfo {author} {\bibfnamefont {C.}~\bibnamefont {Kurtsiefer}}, \bibinfo
  {author} {\bibfnamefont {J.}~\bibnamefont {Skaar}}, \bibinfo {author}
  {\bibfnamefont {V.}~\bibnamefont {Makarov}}, \ and\ \bibinfo {author}
  {\bibfnamefont {I.}~\bibnamefont {Gerhardt}},\ }\href {\doibase
  http://dx.doi.org/10.1063/1.4854615} {\bibfield  {journal} {\bibinfo
  {journal} {Review of Scientific Instruments}\ }\textbf {\bibinfo {volume}
  {85}},\ \bibinfo {eid} {013108} (\bibinfo {year} {2014})}\BibitemShut
  {NoStop}%
\bibitem [{\citenamefont {Tanner}\ \emph {et~al.}(2014)\citenamefont {Tanner},
  \citenamefont {Makarov},\ and\ \citenamefont {Hadfield}}]{tanner14qkdhack}%
  \BibitemOpen
  \bibfield  {author} {\bibinfo {author} {\bibfnamefont {M.~G.}\ \bibnamefont
  {Tanner}}, \bibinfo {author} {\bibfnamefont {V.}~\bibnamefont {Makarov}}, \
  and\ \bibinfo {author} {\bibfnamefont {R.~H.}\ \bibnamefont {Hadfield}},\
  }\href {\doibase 10.1364/OE.22.006734} {\bibfield  {journal} {\bibinfo
  {journal} {Opt. Express}\ }\textbf {\bibinfo {volume} {22}},\ \bibinfo
  {pages} {6734} (\bibinfo {year} {2014})}\BibitemShut {NoStop}%
\bibitem [{\citenamefont {Hall}(2011)}]{hall11relaxed}%
  \BibitemOpen
  \bibfield  {author} {\bibinfo {author} {\bibfnamefont {M.~J.~W.}\
  \bibnamefont {Hall}},\ }\href {\doibase 10.1103/PhysRevA.84.022102}
  {\bibfield  {journal} {\bibinfo  {journal} {Phys. Rev. A}\ }\textbf {\bibinfo
  {volume} {84}},\ \bibinfo {pages} {022102} (\bibinfo {year}
  {2011})}\BibitemShut {NoStop}%
\bibitem [{\citenamefont {Lo}\ \emph {et~al.}(2012)\citenamefont {Lo},
  \citenamefont {Curty},\ and\ \citenamefont {Qi}}]{measdi}%
  \BibitemOpen
  \bibfield  {author} {\bibinfo {author} {\bibfnamefont {H.-K.}\ \bibnamefont
  {Lo}}, \bibinfo {author} {\bibfnamefont {M.}~\bibnamefont {Curty}}, \ and\
  \bibinfo {author} {\bibfnamefont {B.}~\bibnamefont {Qi}},\ }\href@noop {}
  {\bibfield  {journal} {\bibinfo  {journal} {Physical Review Letters}\
  }\textbf {\bibinfo {volume} {108}},\ \bibinfo {pages} {130503} (\bibinfo
  {year} {2012})}\BibitemShut {NoStop}%
\bibitem [{\citenamefont {Pawlowski}\ and\ \citenamefont
  {Brunner}(2011)}]{Marcin2011}%
  \BibitemOpen
  \bibfield  {author} {\bibinfo {author} {\bibfnamefont {M.}~\bibnamefont
  {Pawlowski}}\ and\ \bibinfo {author} {\bibfnamefont {N.}~\bibnamefont
  {Brunner}},\ }\href@noop {} {\bibfield  {journal} {\bibinfo  {journal}
  {Physical Review A}\ }\textbf {\bibinfo {volume} {84}},\ \bibinfo {pages}
  {010203(R)} (\bibinfo {year} {2011})}\BibitemShut {NoStop}%
\bibitem [{\citenamefont {Branciard}\ \emph {et~al.}(2012)\citenamefont
  {Branciard}, \citenamefont {Cavalcanti}, \citenamefont {Walborn},
  \citenamefont {Scarani},\ and\ \citenamefont {Wiseman}}]{steer12}%
  \BibitemOpen
  \bibfield  {author} {\bibinfo {author} {\bibfnamefont {C.}~\bibnamefont
  {Branciard}}, \bibinfo {author} {\bibfnamefont {E.~G.}\ \bibnamefont
  {Cavalcanti}}, \bibinfo {author} {\bibfnamefont {S.~P.}\ \bibnamefont
  {Walborn}}, \bibinfo {author} {\bibfnamefont {V.}~\bibnamefont {Scarani}}, \
  and\ \bibinfo {author} {\bibfnamefont {H.}~\bibnamefont {Wiseman}},\
  }\href@noop {} {\bibfield  {journal} {\bibinfo  {journal} {Physical Review
  A}\ }\textbf {\bibinfo {volume} {85}},\ \bibinfo {pages} {010301(R)}
  (\bibinfo {year} {2012})}\BibitemShut {NoStop}%
\bibitem [{\citenamefont {Bancal}\ \emph {et~al.}(2014)\citenamefont {Bancal},
  \citenamefont {Sheridan},\ and\ \citenamefont {Scarani}}]{bancal13more}%
  \BibitemOpen
  \bibfield  {author} {\bibinfo {author} {\bibfnamefont {J.-D.}\ \bibnamefont
  {Bancal}}, \bibinfo {author} {\bibfnamefont {L.}~\bibnamefont {Sheridan}}, \
  and\ \bibinfo {author} {\bibfnamefont {V.}~\bibnamefont {Scarani}},\ }\href
  {http://stacks.iop.org/1367-2630/16/i=3/a=033011} {\bibfield  {journal}
  {\bibinfo  {journal} {New Journal of Physics}\ }\textbf {\bibinfo {volume}
  {16}},\ \bibinfo {pages} {033011} (\bibinfo {year} {2014})}\BibitemShut
  {NoStop}%
\bibitem [{\citenamefont {{Giustina}}\ \emph {et~al.}(2013)\citenamefont
  {{Giustina}}, \citenamefont {{Mech}}, \citenamefont {{Ramelow}},
  \citenamefont {{Wittmann}}, \citenamefont {{Kofler}}, \citenamefont
  {{Beyer}}, \citenamefont {{Lita}}, \citenamefont {{Calkins}}, \citenamefont
  {{Gerrits}}, \citenamefont {{Nam}}, \citenamefont {{Ursin}},\ and\
  \citenamefont {{Zeilinger}}}]{giustina12}%
  \BibitemOpen
  \bibfield  {author} {\bibinfo {author} {\bibfnamefont {M.}~\bibnamefont
  {{Giustina}}}, \bibinfo {author} {\bibfnamefont {A.}~\bibnamefont {{Mech}}},
  \bibinfo {author} {\bibfnamefont {S.}~\bibnamefont {{Ramelow}}}, \bibinfo
  {author} {\bibfnamefont {B.}~\bibnamefont {{Wittmann}}}, \bibinfo {author}
  {\bibfnamefont {J.}~\bibnamefont {{Kofler}}}, \bibinfo {author}
  {\bibfnamefont {J.}~\bibnamefont {{Beyer}}}, \bibinfo {author} {\bibfnamefont
  {A.}~\bibnamefont {{Lita}}}, \bibinfo {author} {\bibfnamefont
  {B.}~\bibnamefont {{Calkins}}}, \bibinfo {author} {\bibfnamefont
  {T.}~\bibnamefont {{Gerrits}}}, \bibinfo {author} {\bibfnamefont {S.~W.}\
  \bibnamefont {{Nam}}}, \bibinfo {author} {\bibfnamefont {R.}~\bibnamefont
  {{Ursin}}}, \ and\ \bibinfo {author} {\bibfnamefont {A.}~\bibnamefont
  {{Zeilinger}}},\ }\href {\doibase 10.1038/nature12012} {\bibfield  {journal}
  {\bibinfo  {journal} {\nat}\ }\textbf {\bibinfo {volume} {497}},\ \bibinfo
  {pages} {227} (\bibinfo {year} {2013})},\ \Eprint
  {http://arxiv.org/abs/1212.0533} {arXiv:1212.0533 [quant-ph]} \BibitemShut
  {NoStop}%
\bibitem [{\citenamefont {Christensen}\ \emph {et~al.}(2013)\citenamefont
  {Christensen}, \citenamefont {McCusker}, \citenamefont {Altepeter},
  \citenamefont {Calkins}, \citenamefont {Gerrits}, \citenamefont {Lita},
  \citenamefont {Miller}, \citenamefont {Shalm}, \citenamefont {Zhang},
  \citenamefont {Nam}, \citenamefont {Brunner}, \citenamefont {Lim},
  \citenamefont {Gisin},\ and\ \citenamefont {Kwiat}}]{kwiat13exp}%
  \BibitemOpen
  \bibfield  {author} {\bibinfo {author} {\bibfnamefont {B.~G.}\ \bibnamefont
  {Christensen}}, \bibinfo {author} {\bibfnamefont {K.~T.}\ \bibnamefont
  {McCusker}}, \bibinfo {author} {\bibfnamefont {J.~B.}\ \bibnamefont
  {Altepeter}}, \bibinfo {author} {\bibfnamefont {B.}~\bibnamefont {Calkins}},
  \bibinfo {author} {\bibfnamefont {T.}~\bibnamefont {Gerrits}}, \bibinfo
  {author} {\bibfnamefont {A.~E.}\ \bibnamefont {Lita}}, \bibinfo {author}
  {\bibfnamefont {A.}~\bibnamefont {Miller}}, \bibinfo {author} {\bibfnamefont
  {L.~K.}\ \bibnamefont {Shalm}}, \bibinfo {author} {\bibfnamefont
  {Y.}~\bibnamefont {Zhang}}, \bibinfo {author} {\bibfnamefont {S.~W.}\
  \bibnamefont {Nam}}, \bibinfo {author} {\bibfnamefont {N.}~\bibnamefont
  {Brunner}}, \bibinfo {author} {\bibfnamefont {C.~C.~W.}\ \bibnamefont {Lim}},
  \bibinfo {author} {\bibfnamefont {N.}~\bibnamefont {Gisin}}, \ and\ \bibinfo
  {author} {\bibfnamefont {P.~G.}\ \bibnamefont {Kwiat}},\ }\href {\doibase
  10.1103/PhysRevLett.111.130406} {\bibfield  {journal} {\bibinfo  {journal}
  {Phys. Rev. Lett.}\ }\textbf {\bibinfo {volume} {111}},\ \bibinfo {pages}
  {130406} (\bibinfo {year} {2013})}\BibitemShut {NoStop}%
\bibitem [{\citenamefont {{Gill}}(2003)}]{gill03}%
  \BibitemOpen
  \bibfield  {author} {\bibinfo {author} {\bibfnamefont {R.~D.}\ \bibnamefont
  {{Gill}}},\ }\href@noop {} {\bibfield  {journal} {\bibinfo  {journal} {eprint
  arXiv:quant-ph/0301059}\ } (\bibinfo {year} {2003})},\ \Eprint
  {http://arxiv.org/abs/quant-ph/0301059} {quant-ph/0301059} \BibitemShut
  {NoStop}%
\bibitem [{\citenamefont {Zhang}\ \emph {et~al.}(2011)\citenamefont {Zhang},
  \citenamefont {Glancy},\ and\ \citenamefont {Knill}}]{knill11}%
  \BibitemOpen
  \bibfield  {author} {\bibinfo {author} {\bibfnamefont {Y.}~\bibnamefont
  {Zhang}}, \bibinfo {author} {\bibfnamefont {S.}~\bibnamefont {Glancy}}, \
  and\ \bibinfo {author} {\bibfnamefont {E.}~\bibnamefont {Knill}},\ }\href
  {\doibase 10.1103/PhysRevA.84.062118} {\bibfield  {journal} {\bibinfo
  {journal} {Phys. Rev. A}\ }\textbf {\bibinfo {volume} {84}},\ \bibinfo
  {pages} {062118} (\bibinfo {year} {2011})}\BibitemShut {NoStop}%
\bibitem [{\citenamefont {Ac{\'\i}n}\ \emph {et~al.}(2012)\citenamefont
  {Ac{\'\i}n}, \citenamefont {Massar},\ and\ \citenamefont
  {Pironio}}]{acin12randomness}%
  \BibitemOpen
  \bibfield  {author} {\bibinfo {author} {\bibfnamefont {A.}~\bibnamefont
  {Ac{\'\i}n}}, \bibinfo {author} {\bibfnamefont {S.}~\bibnamefont {Massar}}, \
  and\ \bibinfo {author} {\bibfnamefont {S.}~\bibnamefont {Pironio}},\
  }\href@noop {} {\bibfield  {journal} {\bibinfo  {journal} {Physical Review
  Letters}\ }\textbf {\bibinfo {volume} {108}},\ \bibinfo {pages} {100402}
  (\bibinfo {year} {2012})}\BibitemShut {NoStop}%
\bibitem [{foo()}]{footnote}%
  \BibitemOpen
  \href@noop {} {\bibinfo  {journal} {This is the case for instance in the
  tomography level of trust addressed below, since the optimal guessing
  strategy for different settings is always compatible with the known
  measurements performed on the purification of the observed state. In the
  device-independent level of trust, however, the optimal guessing
  probabilities for two different settings may in principle require the
  realization of different states and measurements. In this case, more settings
  could help as well. We thank an anonymous referee for bringing this point to
  our attention.}\ }\BibitemShut {NoStop}%
\bibitem [{\citenamefont {Nieto-Silleras}\ \emph {et~al.}(2014)\citenamefont
  {Nieto-Silleras}, \citenamefont {Pironio},\ and\ \citenamefont
  {Silman}}]{silleras14more}%
  \BibitemOpen
\bibfield  {journal} {  }\bibfield  {author} {\bibinfo {author} {\bibfnamefont
  {O.}~\bibnamefont {Nieto-Silleras}}, \bibinfo {author} {\bibfnamefont
  {S.}~\bibnamefont {Pironio}}, \ and\ \bibinfo {author} {\bibfnamefont
  {J.}~\bibnamefont {Silman}},\ }\href
  {http://stacks.iop.org/1367-2630/16/i=1/a=013035} {\bibfield  {journal}
  {\bibinfo  {journal} {New Journal of Physics}\ }\textbf {\bibinfo {volume}
  {16}},\ \bibinfo {pages} {013035} (\bibinfo {year} {2014})}\BibitemShut
  {NoStop}%
\bibitem [{\citenamefont {Tomamichel}\ and\ \citenamefont
  {Renner}(2011)}]{tomamichel11}%
  \BibitemOpen
  \bibfield  {author} {\bibinfo {author} {\bibfnamefont {M.}~\bibnamefont
  {Tomamichel}}\ and\ \bibinfo {author} {\bibfnamefont {R.}~\bibnamefont
  {Renner}},\ }\href {\doibase 10.1103/PhysRevLett.106.110506} {\bibfield
  {journal} {\bibinfo  {journal} {Phys. Rev. Lett.}\ }\textbf {\bibinfo
  {volume} {106}},\ \bibinfo {pages} {110506} (\bibinfo {year}
  {2011})}\BibitemShut {NoStop}%
\bibitem [{\citenamefont {Jones}\ \emph {et~al.}(2007)\citenamefont {Jones},
  \citenamefont {Wiseman},\ and\ \citenamefont
  {Doherty}}]{jones2007entanglement}%
  \BibitemOpen
  \bibfield  {author} {\bibinfo {author} {\bibfnamefont {S.~J.}\ \bibnamefont
  {Jones}}, \bibinfo {author} {\bibfnamefont {H.~M.}\ \bibnamefont {Wiseman}},
  \ and\ \bibinfo {author} {\bibfnamefont {A.~C.}\ \bibnamefont {Doherty}},\
  }\href@noop {} {\bibfield  {journal} {\bibinfo  {journal} {Physical Review
  A}\ }\textbf {\bibinfo {volume} {76}},\ \bibinfo {pages} {052116} (\bibinfo
  {year} {2007})}\BibitemShut {NoStop}%
\bibitem [{\citenamefont {Cavalcanti}\ \emph {et~al.}(2009)\citenamefont
  {Cavalcanti}, \citenamefont {Jones}, \citenamefont {Wiseman},\ and\
  \citenamefont {Reid}}]{cavalcanti2009experimental}%
  \BibitemOpen
  \bibfield  {author} {\bibinfo {author} {\bibfnamefont {E.~G.}\ \bibnamefont
  {Cavalcanti}}, \bibinfo {author} {\bibfnamefont {S.~J.}\ \bibnamefont
  {Jones}}, \bibinfo {author} {\bibfnamefont {H.~M.}\ \bibnamefont {Wiseman}},
  \ and\ \bibinfo {author} {\bibfnamefont {M.}~\bibnamefont {Reid}},\
  }\href@noop {} {\bibfield  {journal} {\bibinfo  {journal} {Physical Review
  A}\ }\textbf {\bibinfo {volume} {80}},\ \bibinfo {pages} {032112} (\bibinfo
  {year} {2009})}\BibitemShut {NoStop}%
\bibitem [{\citenamefont {Navascu{\'e}s}\ \emph {et~al.}(2008)\citenamefont
  {Navascu{\'e}s}, \citenamefont {Pironio},\ and\ \citenamefont
  {Ac{\'\i}n}}]{qtest}%
  \BibitemOpen
  \bibfield  {author} {\bibinfo {author} {\bibfnamefont {M.}~\bibnamefont
  {Navascu{\'e}s}}, \bibinfo {author} {\bibfnamefont {S.}~\bibnamefont
  {Pironio}}, \ and\ \bibinfo {author} {\bibfnamefont {A.}~\bibnamefont
  {Ac{\'\i}n}},\ }\href@noop {} {\bibfield  {journal} {\bibinfo  {journal} {New
  J. Phys.}\ }\textbf {\bibinfo {volume} {10}},\ \bibinfo {pages} {073013}
  (\bibinfo {year} {2008})}\BibitemShut {NoStop}%
\bibitem [{\citenamefont {Popescu}\ and\ \citenamefont
  {Rohrlich}(1992)}]{popescu92states}%
  \BibitemOpen
  \bibfield  {author} {\bibinfo {author} {\bibfnamefont {S.}~\bibnamefont
  {Popescu}}\ and\ \bibinfo {author} {\bibfnamefont {D.}~\bibnamefont
  {Rohrlich}},\ }\href@noop {} {\bibfield  {journal} {\bibinfo  {journal}
  {Physics Letters A}\ }\textbf {\bibinfo {volume} {169}},\ \bibinfo {pages}
  {411} (\bibinfo {year} {1992})}\BibitemShut {NoStop}%
\bibitem [{\citenamefont {Saunders}\ \emph {et~al.}(2010)\citenamefont
  {Saunders}, \citenamefont {Jones}, \citenamefont {Wiseman},\ and\
  \citenamefont {Pryde}}]{saunders10exp}%
  \BibitemOpen
  \bibfield  {author} {\bibinfo {author} {\bibfnamefont {D.~J.}\ \bibnamefont
  {Saunders}}, \bibinfo {author} {\bibfnamefont {S.~J.}\ \bibnamefont {Jones}},
  \bibinfo {author} {\bibfnamefont {H.~M.}\ \bibnamefont {Wiseman}}, \ and\
  \bibinfo {author} {\bibfnamefont {G.}~\bibnamefont {Pryde}},\ }\href@noop {}
  {\bibfield  {journal} {\bibinfo  {journal} {Nature Physics}\ }\textbf
  {\bibinfo {volume} {6}},\ \bibinfo {pages} {845} (\bibinfo {year}
  {2010})}\BibitemShut {NoStop}%
\bibitem [{\citenamefont {Maassen}\ and\ \citenamefont
  {Uffink}(1988)}]{maassenuffink}%
  \BibitemOpen
  \bibfield  {author} {\bibinfo {author} {\bibfnamefont {H.}~\bibnamefont
  {Maassen}}\ and\ \bibinfo {author} {\bibfnamefont {J.}~\bibnamefont
  {Uffink}},\ }\href@noop {} {\bibfield  {journal} {\bibinfo  {journal} {Phys.
  Rev. Lett.}\ }\textbf {\bibinfo {volume} {60}},\ \bibinfo {pages} {1103}
  (\bibinfo {year} {1988})}\BibitemShut {NoStop}%
\bibitem [{\citenamefont {Chen}\ \emph {et~al.}(2007)\citenamefont {Chen},
  \citenamefont {Bergou}, \citenamefont {Zhu},\ and\ \citenamefont
  {Guo}}]{chen07povm}%
  \BibitemOpen
  \bibfield  {author} {\bibinfo {author} {\bibfnamefont {P.-X.}\ \bibnamefont
  {Chen}}, \bibinfo {author} {\bibfnamefont {J.~A.}\ \bibnamefont {Bergou}},
  \bibinfo {author} {\bibfnamefont {S.-Y.}\ \bibnamefont {Zhu}}, \ and\
  \bibinfo {author} {\bibfnamefont {G.-C.}\ \bibnamefont {Guo}},\ }\href
  {\doibase 10.1103/PhysRevA.76.060303} {\bibfield  {journal} {\bibinfo
  {journal} {Phys. Rev. A}\ }\textbf {\bibinfo {volume} {76}},\ \bibinfo
  {pages} {060303} (\bibinfo {year} {2007})}\BibitemShut {NoStop}%
\bibitem [{\citenamefont {Fiorentino}\ \emph {et~al.}(2007)\citenamefont
  {Fiorentino}, \citenamefont {Santori}, \citenamefont {Spillane},
  \citenamefont {Beausoleil},\ and\ \citenamefont {Munro}}]{fiorentino07}%
  \BibitemOpen
  \bibfield  {author} {\bibinfo {author} {\bibfnamefont {M.}~\bibnamefont
  {Fiorentino}}, \bibinfo {author} {\bibfnamefont {C.}~\bibnamefont {Santori}},
  \bibinfo {author} {\bibfnamefont {S.}~\bibnamefont {Spillane}}, \bibinfo
  {author} {\bibfnamefont {R.}~\bibnamefont {Beausoleil}}, \ and\ \bibinfo
  {author} {\bibfnamefont {W.}~\bibnamefont {Munro}},\ }\href@noop {}
  {\bibfield  {journal} {\bibinfo  {journal} {Phys. Rev. A}\ }\textbf {\bibinfo
  {volume} {75}},\ \bibinfo {pages} {032334} (\bibinfo {year}
  {2007})}\BibitemShut {NoStop}%
\bibitem [{\citenamefont {Vallone}\ \emph {et~al.}(2014)\citenamefont
  {Vallone}, \citenamefont {Marangon}, \citenamefont {Tomasin},\ and\
  \citenamefont {Villoresi}}]{vallone}%
  \BibitemOpen
  \bibfield  {author} {\bibinfo {author} {\bibfnamefont {G.}~\bibnamefont
  {Vallone}}, \bibinfo {author} {\bibfnamefont {D.}~\bibnamefont {Marangon}},
  \bibinfo {author} {\bibfnamefont {M.}~\bibnamefont {Tomasin}}, \ and\
  \bibinfo {author} {\bibfnamefont {P.}~\bibnamefont {Villoresi}},\ }\href@noop
  {} {\bibfield  {journal} {\bibinfo  {journal} {arXiv preprint
  arXiv:1401.7917}\ } (\bibinfo {year} {2014})}\BibitemShut {NoStop}%
\bibitem [{\citenamefont {Jennewein}\ \emph {et~al.}(2000)\citenamefont
  {Jennewein}, \citenamefont {Achleitner}, \citenamefont {Weihs}, \citenamefont
  {Weinfurter},\ and\ \citenamefont {Zeilinger}}]{jennewein00}%
  \BibitemOpen
  \bibfield  {author} {\bibinfo {author} {\bibfnamefont {T.}~\bibnamefont
  {Jennewein}}, \bibinfo {author} {\bibfnamefont {U.}~\bibnamefont
  {Achleitner}}, \bibinfo {author} {\bibfnamefont {G.}~\bibnamefont {Weihs}},
  \bibinfo {author} {\bibfnamefont {H.}~\bibnamefont {Weinfurter}}, \ and\
  \bibinfo {author} {\bibfnamefont {A.}~\bibnamefont {Zeilinger}},\ }\href@noop
  {} {\bibfield  {journal} {\bibinfo  {journal} {Rev. Sci. Instr.}\ }\textbf
  {\bibinfo {volume} {71}},\ \bibinfo {pages} {1675} (\bibinfo {year}
  {2000})}\BibitemShut {NoStop}%
\bibitem [{\citenamefont {Stefanov}\ \emph {et~al.}(2000)\citenamefont
  {Stefanov}, \citenamefont {Gisin}, \citenamefont {Guinnard}, \citenamefont
  {Guinnard},\ and\ \citenamefont {Zbinden}}]{stefanov00}%
  \BibitemOpen
  \bibfield  {author} {\bibinfo {author} {\bibfnamefont {A.}~\bibnamefont
  {Stefanov}}, \bibinfo {author} {\bibfnamefont {N.}~\bibnamefont {Gisin}},
  \bibinfo {author} {\bibfnamefont {O.}~\bibnamefont {Guinnard}}, \bibinfo
  {author} {\bibfnamefont {L.}~\bibnamefont {Guinnard}}, \ and\ \bibinfo
  {author} {\bibfnamefont {H.}~\bibnamefont {Zbinden}},\ }\href@noop {}
  {\bibfield  {journal} {\bibinfo  {journal} {J. Mod. Opt.}\ }\textbf {\bibinfo
  {volume} {47}},\ \bibinfo {pages} {595} (\bibinfo {year} {2000})}\BibitemShut
  {NoStop}%
\bibitem [{\citenamefont {Symul}\ \emph {et~al.}(2011)\citenamefont {Symul},
  \citenamefont {Assad},\ and\ \citenamefont {Lam}}]{symul11}%
  \BibitemOpen
  \bibfield  {author} {\bibinfo {author} {\bibfnamefont {T.}~\bibnamefont
  {Symul}}, \bibinfo {author} {\bibfnamefont {S.}~\bibnamefont {Assad}}, \ and\
  \bibinfo {author} {\bibfnamefont {P.}~\bibnamefont {Lam}},\ }\href@noop {}
  {\bibfield  {journal} {\bibinfo  {journal} {Appl. Phys. Lett.}\ }\textbf
  {\bibinfo {volume} {98}},\ \bibinfo {pages} {231103} (\bibinfo {year}
  {2011})}\BibitemShut {NoStop}%
\end{thebibliography}%

\end{document}